\documentclass[amsmath,showpacs,aps,prb,preprint,floatfix,superscriptaddress]{revtex4-1}
\usepackage{bm}
\usepackage{graphics}
\usepackage{graphicx}
\usepackage{amsthm}
\usepackage{amsmath}
\usepackage{wasysym}
\usepackage{amssymb}
\usepackage{enumerate}
\usepackage[dvips]{epsfig}
\usepackage{multirow}

\begin{document}
\title{Renormalization Group Analysis of a Fermionic Hot Spot Model}
\author{Seth Whitsitt}
\affiliation{Department of Physics, Harvard University, Cambridge, MA 02138, USA}
\author{Subir Sachdev}
\affiliation{Department of Physics, Harvard University, Cambridge, MA 02138, USA}
\affiliation{Perimeter Institute for Theoretical Physics, Waterloo, Ontario N2L 2Y5, Canada}
\date{\today}
\begin{abstract}
We present a renormalization group (RG) analysis of a fermionic ``hot spot'' model of interacting electrons on the square lattice. 
We truncate the Fermi surface excitations to linearly dispersing quasiparticles in 
the vicinity of eight hot spots on the Fermi surface, with each hot spot separated from another by the
wavevector $(\pi, \pi)$. This motivated by the importance of these Fermi surface locations to the onset of antiferromagnetic order; however, we allow
for all possible quartic interactions between the fermions, and also for all possible ordering instabilities.
We compute the RG equations for our model, which depend on whether the hot spots are perfectly nested or not, and relate our results to earlier models. We also compute the RG flow of the relevant order parameters for both Hubbard and $J$, $V$ interactions, and present our results for the dominant instabilities in the nested and non-nested cases. In particular, we find that non-nested hot spots with $J$, $V$ interactions have competing singlet $d_{x^2-y^2}$ superconducting and
$d$-form factor incommensurate density wave instabilities. We also investigate the enhancement of incommensurate density waves near experimentally observed 
wavevectors, and find dominant $d$-form factor enhancement for a range of couplings.
\end{abstract}

\maketitle

\section{Introduction}
\label{sec:intro}

A fruitful approach to the physics of correlated electron systems is to begin with an ordinary Fermi liquid, and to consider the approach
to the onset of antiferromagnetism ({\em i.e.\/} spin density wave (SDW) order). 
It is now becoming clear that before we actually reach the state with SDW order, other interesting instabilities can intervene. So, in a sense, the most
interesting part of the study of SDW quantum critical points is not the critical point per se, but the physics we encounter along our journey towards it
from the Fermi liquid.
This paper will develop a theoretical model for this approach, and apply it to the copper oxide superconductors.

It was noted long ago \cite{scalapino:prb86,scalapino:rmp12} that a spin-singlet $d$-wave superconducting instability can appear in metal
with SDW fluctuations, with the SDW collective mode playing the role of the phonon in conventional BCS theory. More recently, $d$-wave superconductivity
has been observed across the SDW quantum critical point in a Monte Carlo study.\cite{berg:science12}
The appearance of additional instabilities in the vicinity of the SDW quantum critical point 
was noted by Metlitski and Sachdev\cite{metlitski:prb10.2} who found a $d$-form factor incommensurate
density wave.

On the experimental front in the cuprate superconductors, evidence has been accumulating for a density wave instability competing with
superconductivity in the non-La based compounds.
Traces of this order were initially seen as periodic modulations in the density of electronic states around vortices is scanning tunneling 
microscopy.\cite{hoffman:sci02} Later, evidence for the density wave order 
also appeared in STM experiments in zero field.\cite{kapitulnik:03,yazdani:04,davis:07,hudson:08} 
In the modern era, the observation of quantum oscillations\cite{louis0,leboeuf,louis2,harrison,suchitra2,greven,suchitra3} 
has been linked\cite{harrison,AADCSS} to charge order observed in NMR and X-ray scattering experiments.\cite{wu:11,ghiringhelli:12,achkar:12,chang:12,wu:13}
And most recently, STM observations have presented direct phase sensitive evidence for a predominant $d$-form factor of the density wave;\cite{fujita:14}
supporting evidence for such a form factor also appears in X-ray experiments.\cite{comin:14}

The original treatment of Fermi liquid-SDW transitions is due to Hertz.\cite{hertz:prb76} In this method, the fermions at the Fermi surface (FS) are completely integrated out, resulting in an effective action for the SDW order parameter. This action can then be studied using standard RG techniques. While Hertz theory is largely correct for $d\geq 3$, it was shown to fail in $d=2$ by Abanov and Chubukov.\cite{abanov:prl00,abanov:prl04} In particular, they argued that the $d=2$ case should be treated by a spin-fermion model where the SDW order parameter couples to low-energy fermions located at the ``hot spots,'' which are defined to be the points on the FS connected by the SDW ordering wave vector. A field theoretic RG analysis of their theory was presented by
 Metlitski and Sachdev\cite{metlitski:prb10.2}: they found a renormalization of the FS towards perfect nesting, as well as an emergent psuedospin symmetry relating enhanced $d$-wave pairing to an incommensurate $d$-form factor density wave order. The density wave in these computations had a wavevector oriented along
 the $(1, \pm 1)$ directions of the square lattice Brillouin zone, while that in the experiments is along the $(1,0)$, $(0,1)$ directions. 
 A number of studies\cite{sachdev:prl13,sau:14,davis:13,efetov:np13,meier:14,allais:14.1,allais:14.2,wang:14,melikyan:14,tsvelik:14,atkinson:14,metzner,husemann,yamase,kee,bulut} have
 since addressed the physics of the density wave away from close proximity to the SDW critical point, and found parameter regimes where 
 its orientation is along the observed directions.
 
In this paper we consider a purely fermionic analog to the field-theoretic Abanov-Chubukov model in which 
we do not explicitly prefer the interactions associated with SDW ordering.
Instead, we include all possible interactions between the fermions, and (in principle) allow for all instabilities including that of SDW order.
Thus we de-emphasize the role of SDW fluctuations, and retain its memory only in our decision to focus on 
a linearized fermion spectrum in the vicinities of the hot spots.
The resulting theory of interacting hot spot fermions resembles a one-dimensional system, and we use the $g$-ology approach.\cite{solyom,Giamarchi} In this method, we write down all possible quartic couplings between the hot spots, and perform an RG analysis to determine the behavior of the system at low energy. The case where the hot spots are not nested is equivalent to a model studied by Furukawa and Rice,\cite{furukawa98} who found that the model with repulsive Hubbard interactions contained an enhanced incommensurate SDW order. More recently, Carvalho and Freire\cite{freire:npb} investigated the fermionic hot spot model with perfect nesting and claim to find an insulating spin-gapped state with no long-range antiferromagnetic order; their more recent work\cite{freire:aop} appeared while our work was largely complete, and has results related to those presented below. See also the work of Sedeki {\em et al.\/}\cite{bourbonnais:12} on the quasi-one dimensional
case.

We revisit the fermionic hot spot model, extending previous results, and analyze the dominant instabilities of the model in the presence of both Hubbard and $J$, $V$ interactions. In the nested case we find that the model with repulsive Hubbard interactions has dominant N\'eel order, while the presence of $J$, $V$ interactions exhibits enhanced $d_{x^2-y^2}$ pairing for large $J$, crossing over to commensurate charge density wave (CDW) for large $V$. Furthermore, we find that the non-nested case with $J$, $V$ interactions can lead to competition between the $d_{x^2-y^2}$ superconducting order and incommensurate $d$-form factor charge order, in qualitative agreement with experiment, albeit with a diagonal orientation for the ordering wavevector. 
We also look at the enhancement to the charge order with the physically relevant wave vector in Section~\ref{sec:coq}:
while this channel is irrelevant under RG flows, we find an instability at the Hartree-Fock level to charge order with a primarily $d$-form factor component.

\section{Fermionic Hot Spot Model}
\label{sec:model}

\begin{figure}
\includegraphics[width=10cm]{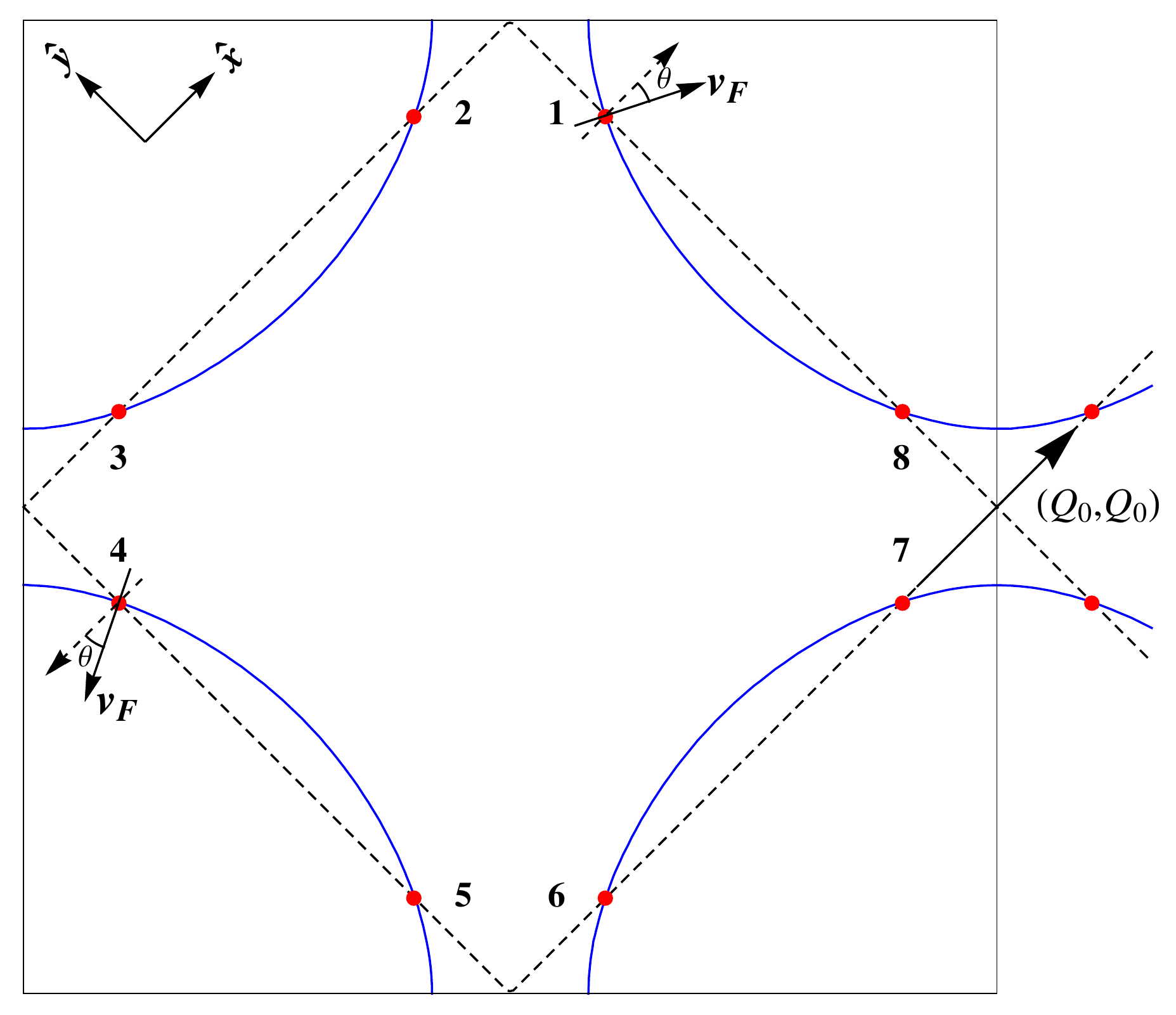}
\caption{(Color online) Fermi surface for the hot spot model with the dispersion relevant to the cuprates. The hot spots are the filled circles, the angle $\theta$ determines whether nesting occurs or not, and the labels on the hot spots are used to define the order parameters in Section \ref{sec:rg}.} 
\label{fig:hotspot}
\end{figure}

We begin our analysis by considering free fermions at zero temperature on a two-dimensional square lattice with first- and second-nearest neighbor hoppings of amplitude $t$ and $t'$ respectively. In units of the lattice constant (which we use for the remainder of the paper), the dispersion is $\epsilon_{\mathbf{k}} = -2 t \left( \cos(k_x) + \cos(k_y) \right) - 4 t' \cos(k_x) \cos(k_y) - \mu$. We consider the locus of points on the FS connected by a wave vector $\mathbf{Q} = (\pi,\pi)$. When $0 < |\mu| < 4 |t'|$, there exist eight such points, denoted ``hot spots,'' which are shown in Fig. \ref{fig:hotspot}. We define our model by describing the low-energy dynamics of the fermions with a linear dispersion at each hot spot: $\epsilon_{\mathbf{k}} \approx \mathbf{v}_{F} \left( \mathbf{k} - \mathbf{k}_{F} \right)$.
In what follows, we will perform calculations in the rotated coordinates $\hat{y} = (\hat{k}_y - \hat{k}_x)/\sqrt{2}$ and $\hat{x} = (\hat{k}_y + \hat{k}_x)/\sqrt{2}$ (see Fig. \ref{fig:hotspot}). Since the approximation of a linear dispersion only holds in a finite region around each hot spot, we introduce a hard momentum cutoff $k_c$ in the $\hat{x}$ and $\hat{y}$ directions centered at each hot spot, and neglect states with momenta larger than these. We also define the energy bandwidth $E_c = 2 v_F k_c$. The cutoff choice is arbitrary, and should not alter the low-energy physics significantly.

We now consider this model in the presence of interactions. Our approach will mirror the ``$g$-ology'' method from one-dimensional physics:\cite{solyom} We add an interacting Hamiltonian to our original model which contains all possible spin-independent interactions which are quartic in the fermions. Our model is described by the action $\mathcal{S} = \mathcal{S}_0 + \mathcal{S}'$ with
\begin{eqnarray}\label{eqn:hamiltonian}
\mathcal{S}_0 &=& \sum_{\sigma} \int \frac{d^3 k}{(2 \pi)^3} \overline{\psi}_{\sigma}(k) \left( \omega - \epsilon_{\mathbf{k}} \right) \psi_{\sigma}(k) \nonumber \\
\mathcal{S}' &=& - \sum_{\sigma \sigma'} \sum_{n}g^B_{n}\int \left( \prod_{i=1}^4 d^3k_i \right) \bar{\delta}(k_1 + k_2 - k_3 - k_4) \overline{\psi}_{\sigma }(k_4) \overline{\psi}_{\sigma' }(k_3) \psi_{\sigma' }(k_2) \psi_{\sigma}(k_1) 
\end{eqnarray}
where we have defined $k_i = (\omega_i,\mathbf{k}_i)$ and $\int d^3k_i = \int d\omega_i \int d^2\mathbf{k}_i$ for each three-momentum, and the modified delta function $\bar{\delta}$ conserves momentum modulo an Umklapp vector. The superscript on the $g^B$ indicates that these are the bare couplings, distinct from the renormalized couplings defined later.

A related model has already been treated within the regime $\theta \in (0,\pi/4)$.\cite{furukawa98} While this appears to be the relevant case for the cuprates, there is reason to investigate the $\theta = 0$ case. Field-theoretic RG studies of the related hot spot models have found that $\theta$ renormalizes to zero.\cite{abanov:prl00,metlitski:prb10.2,sungsik} We will see below that $\theta$ does not renormalize at one-loop in the present model, which might be a sign that the present model ignores the dynamic nesting observed in the previous studies. We would like to investigate the possible effects of nesting, we consider both $\theta = 0$ and $\theta >0$. We note that the precise value of $\theta$ has no effect on the RG behavior of the model in the absence of nesting.

\begin{figure}[h]
\includegraphics[height=3.25cm]{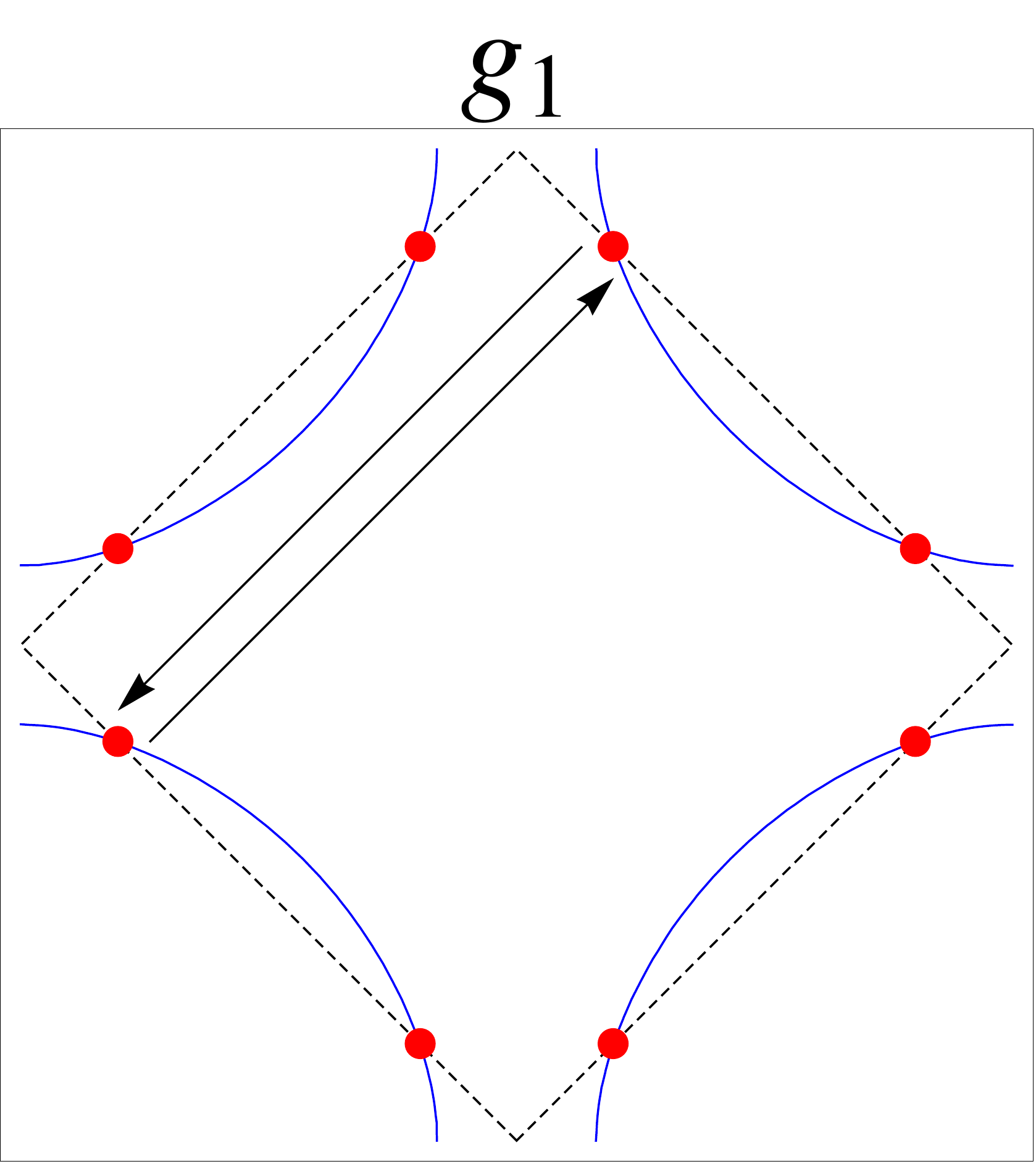}
\includegraphics[height=3.25cm]{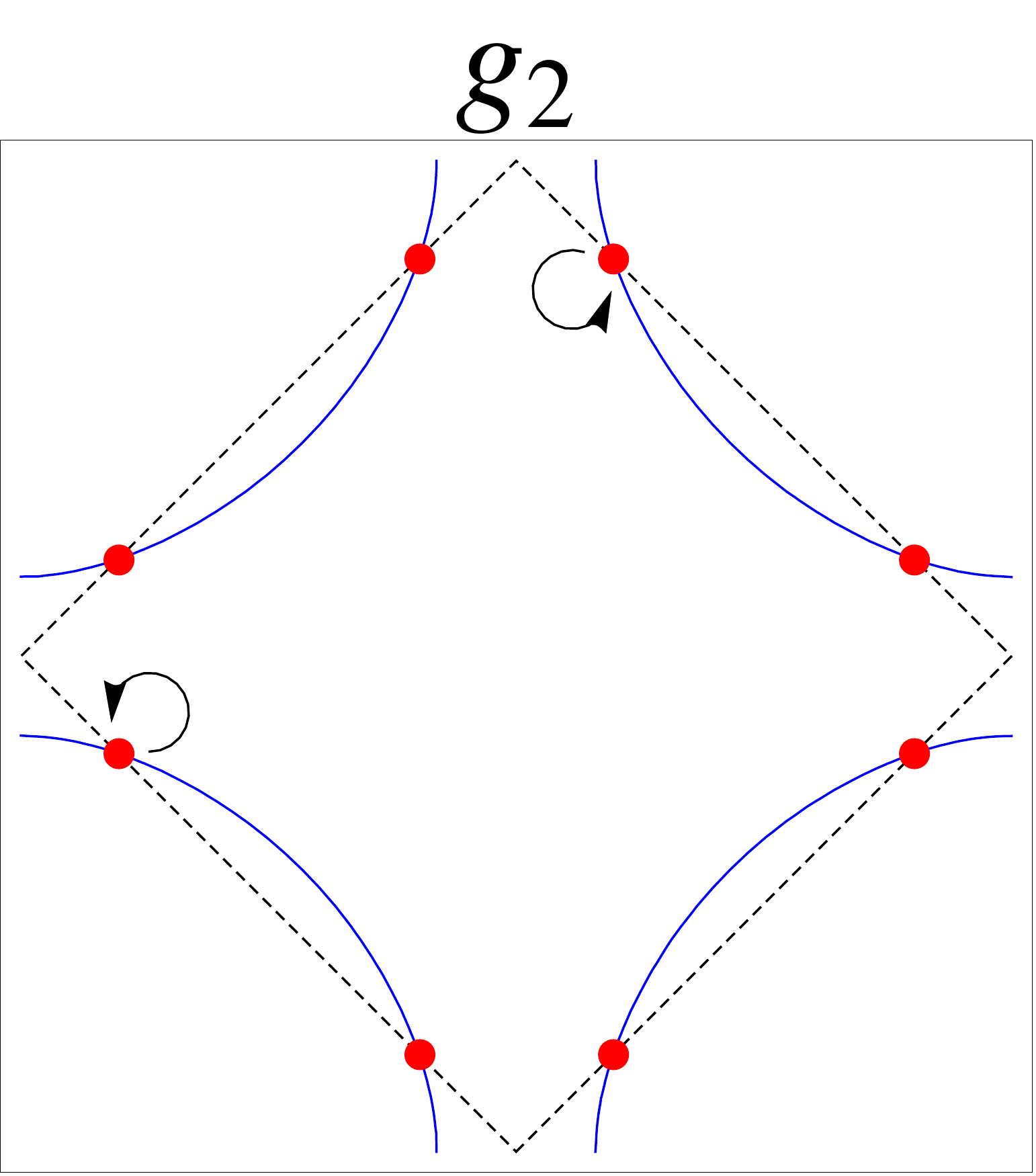}
\includegraphics[height=3.25cm]{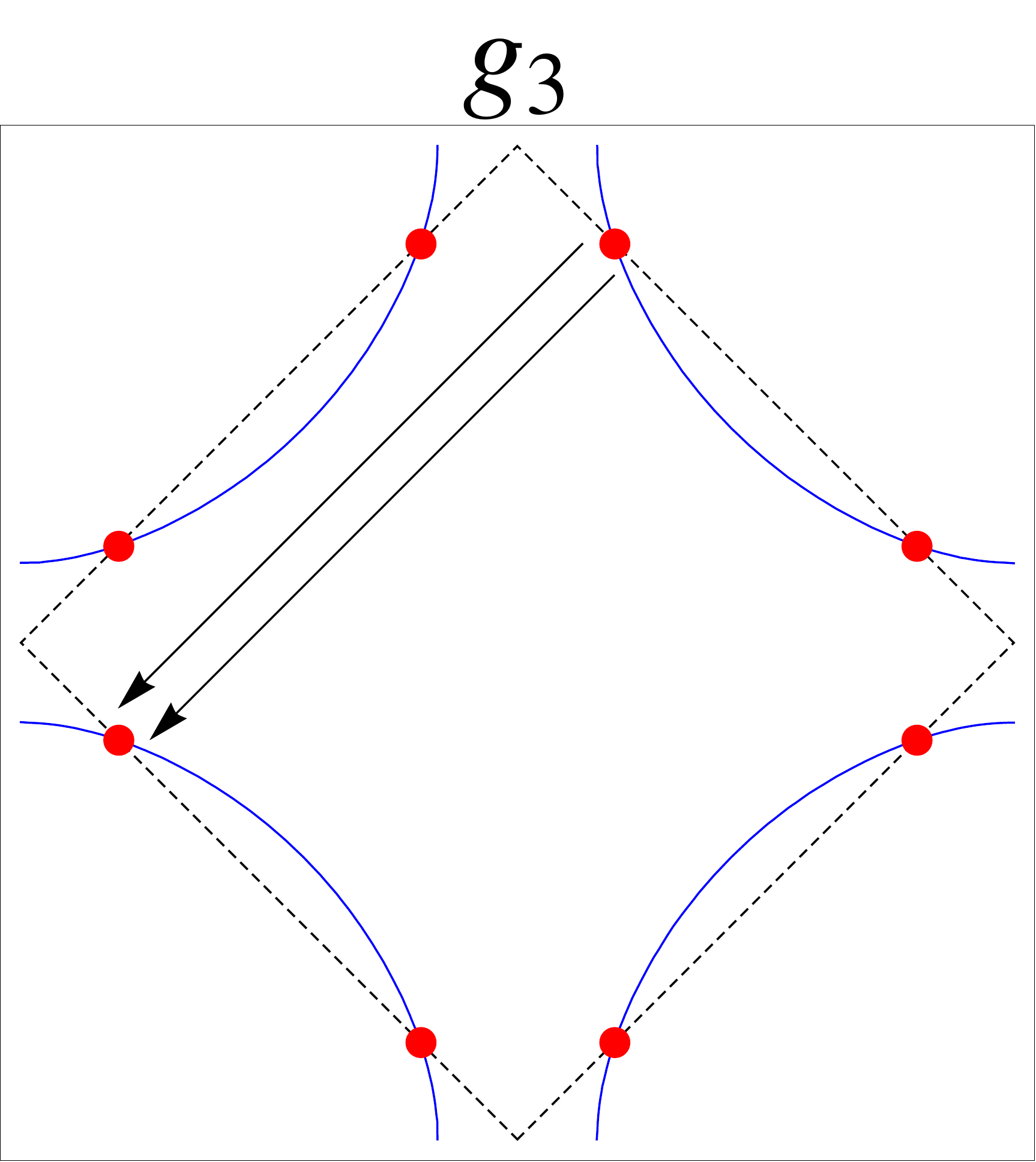}
\includegraphics[height=3.25cm]{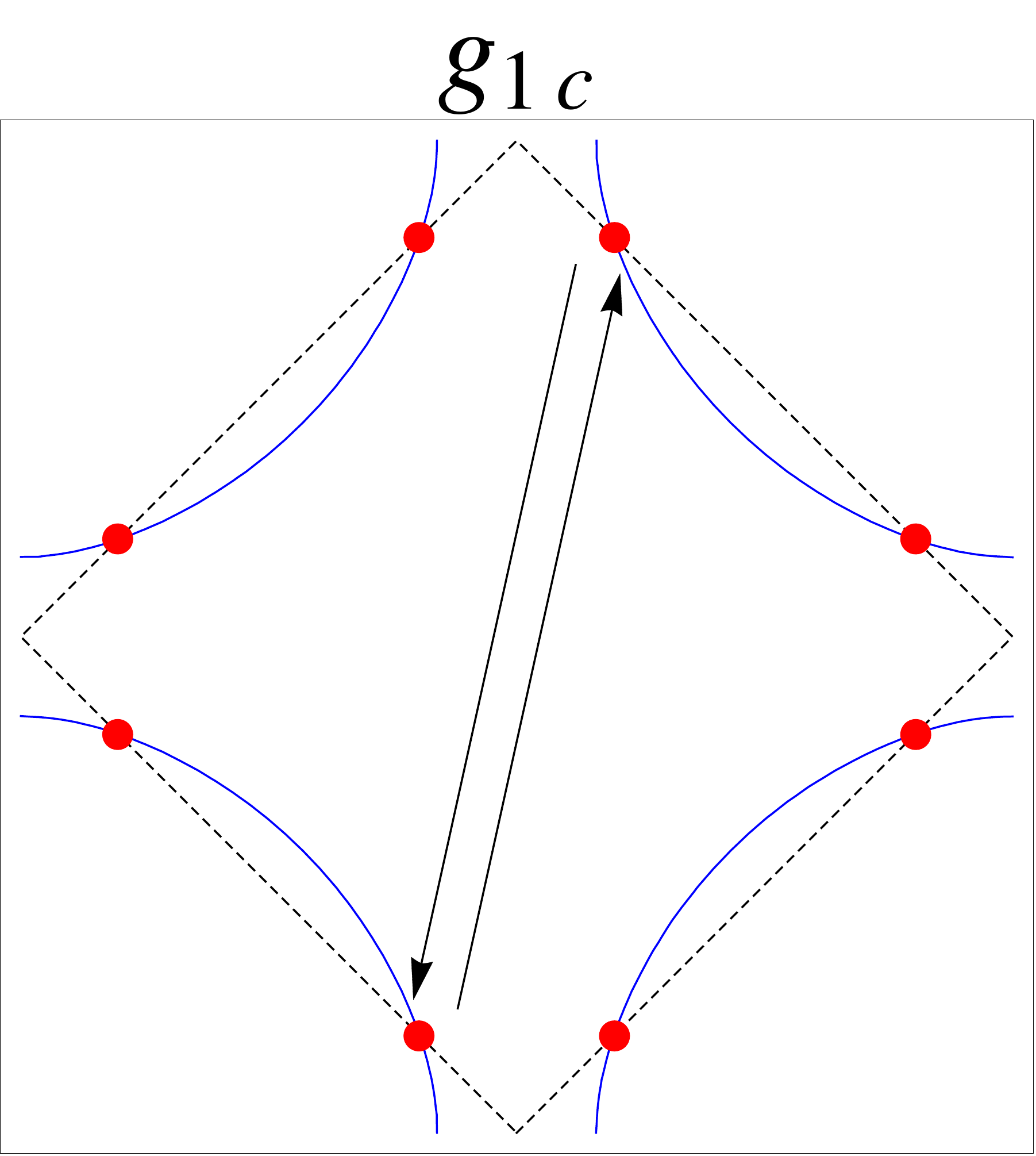}
\includegraphics[height=3.25cm]{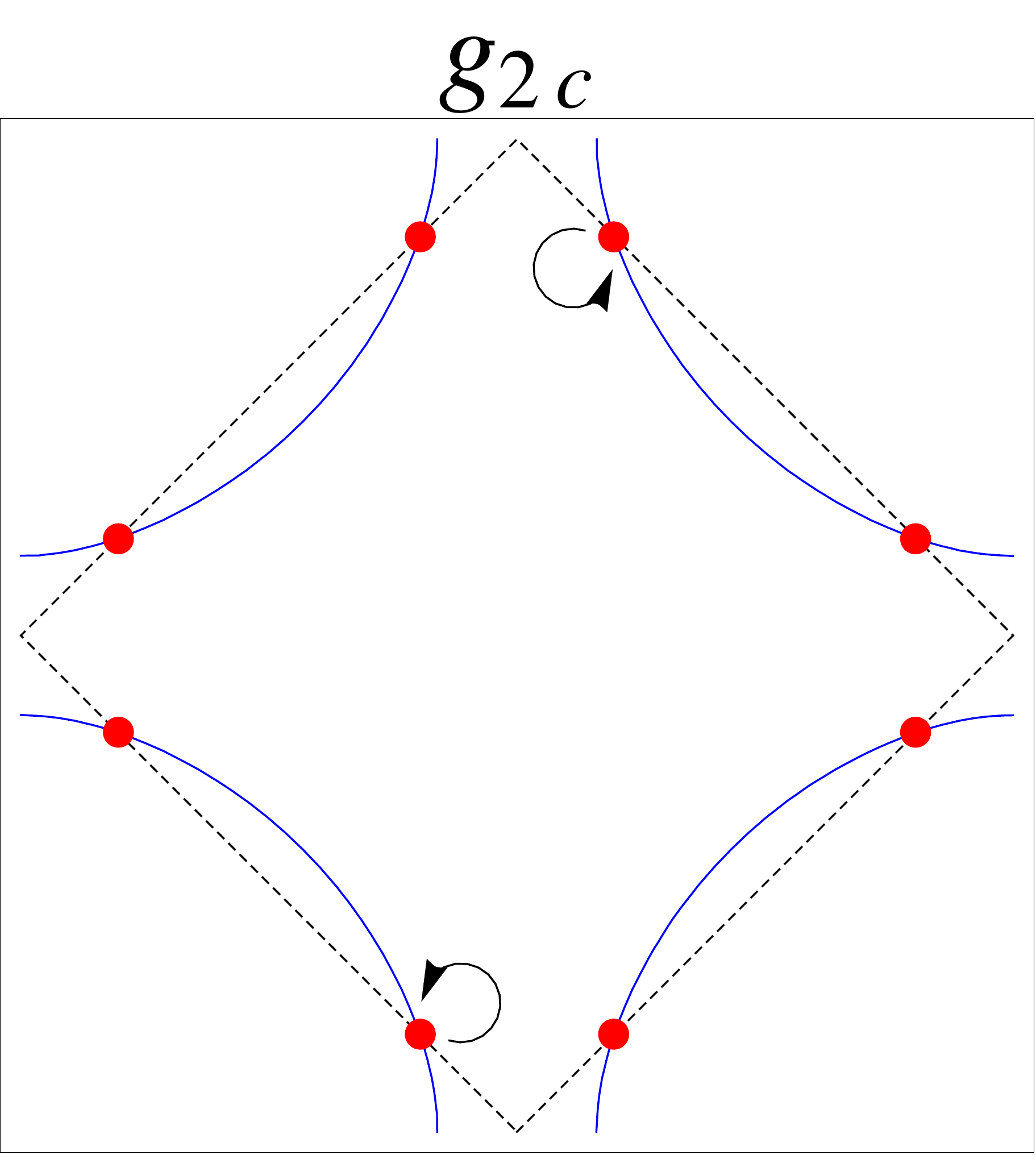}
\includegraphics[height=3.25cm]{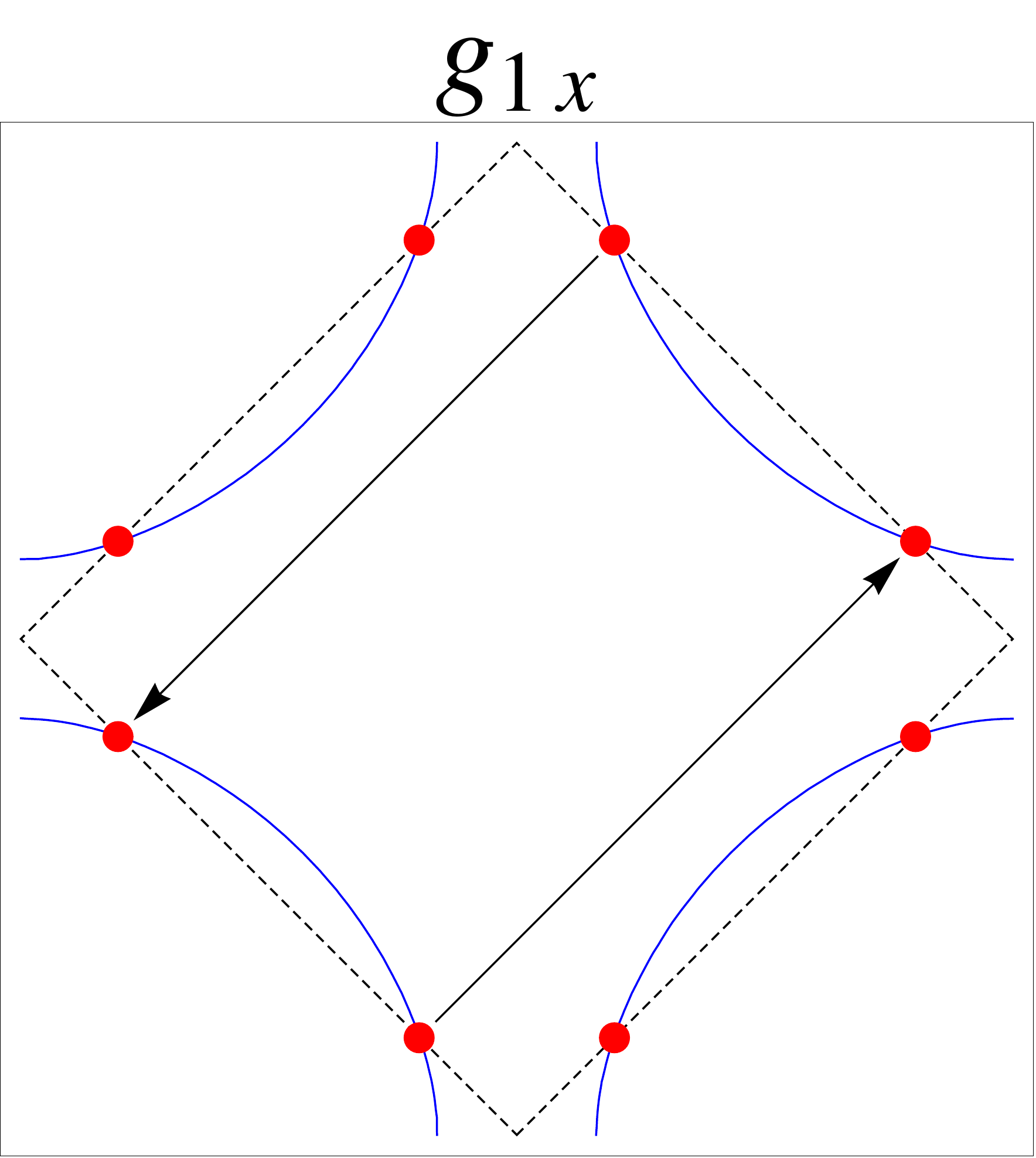}
\includegraphics[height=3.25cm]{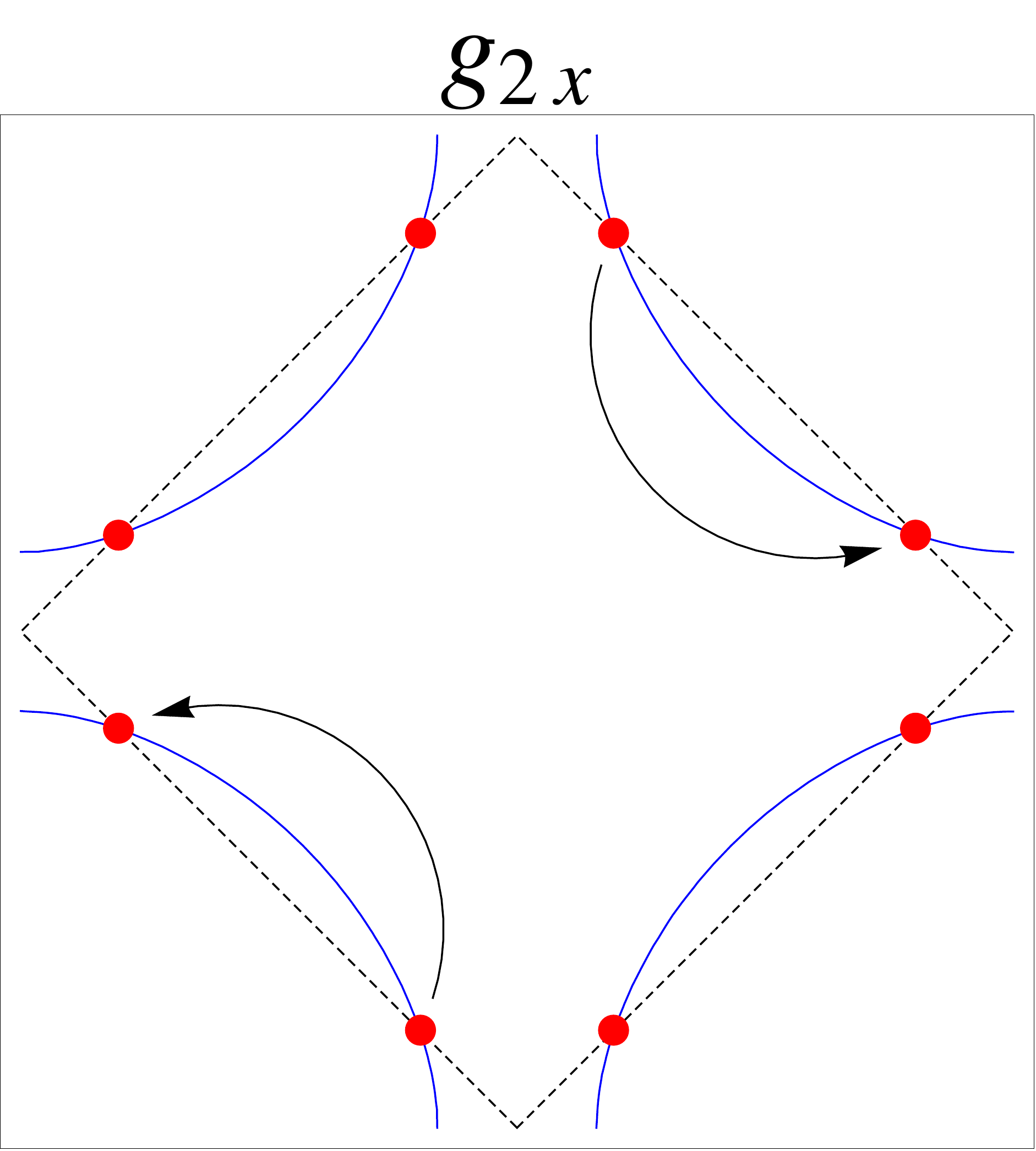}
\includegraphics[height=3.25cm]{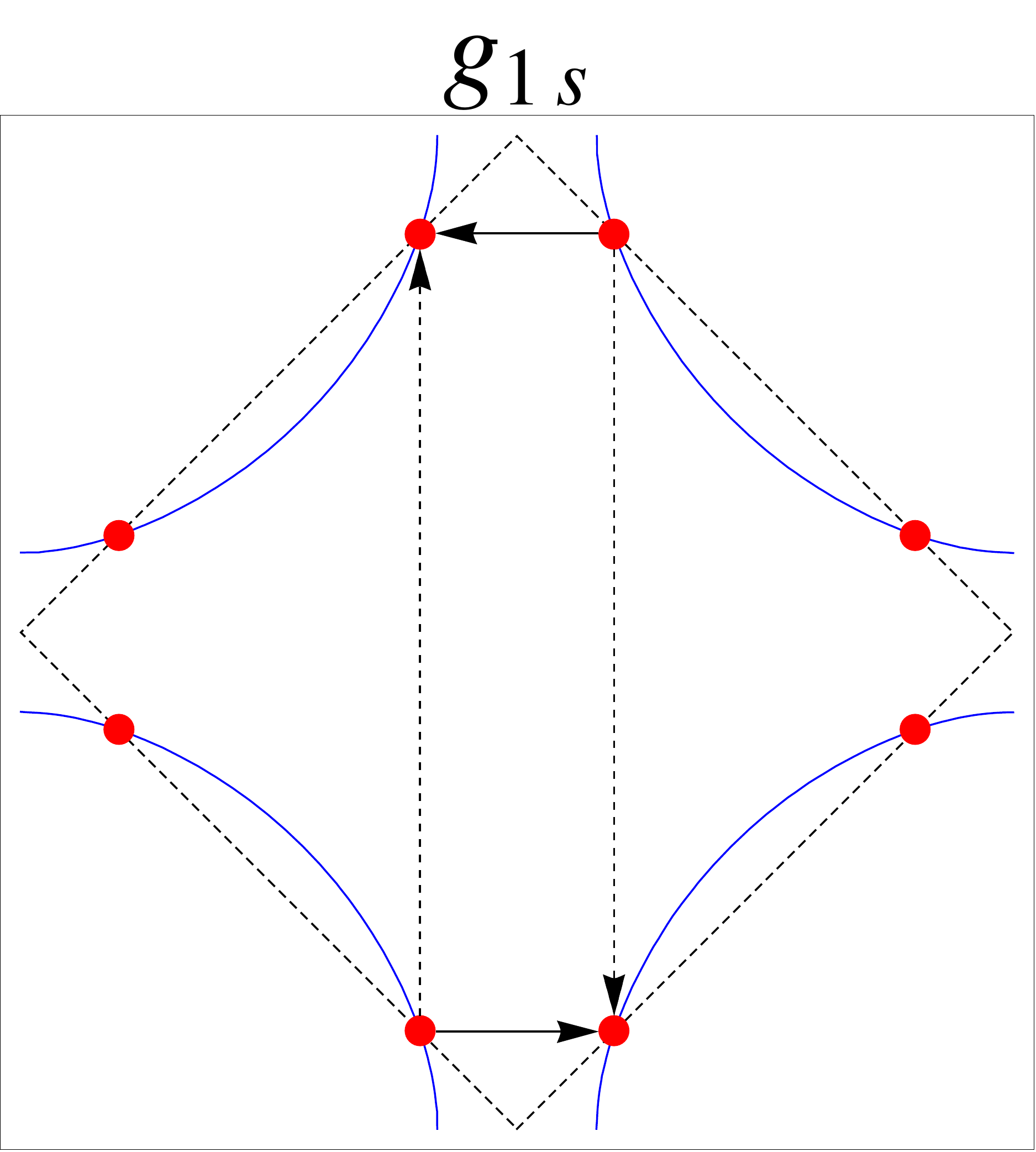}
\includegraphics[height=3.25cm]{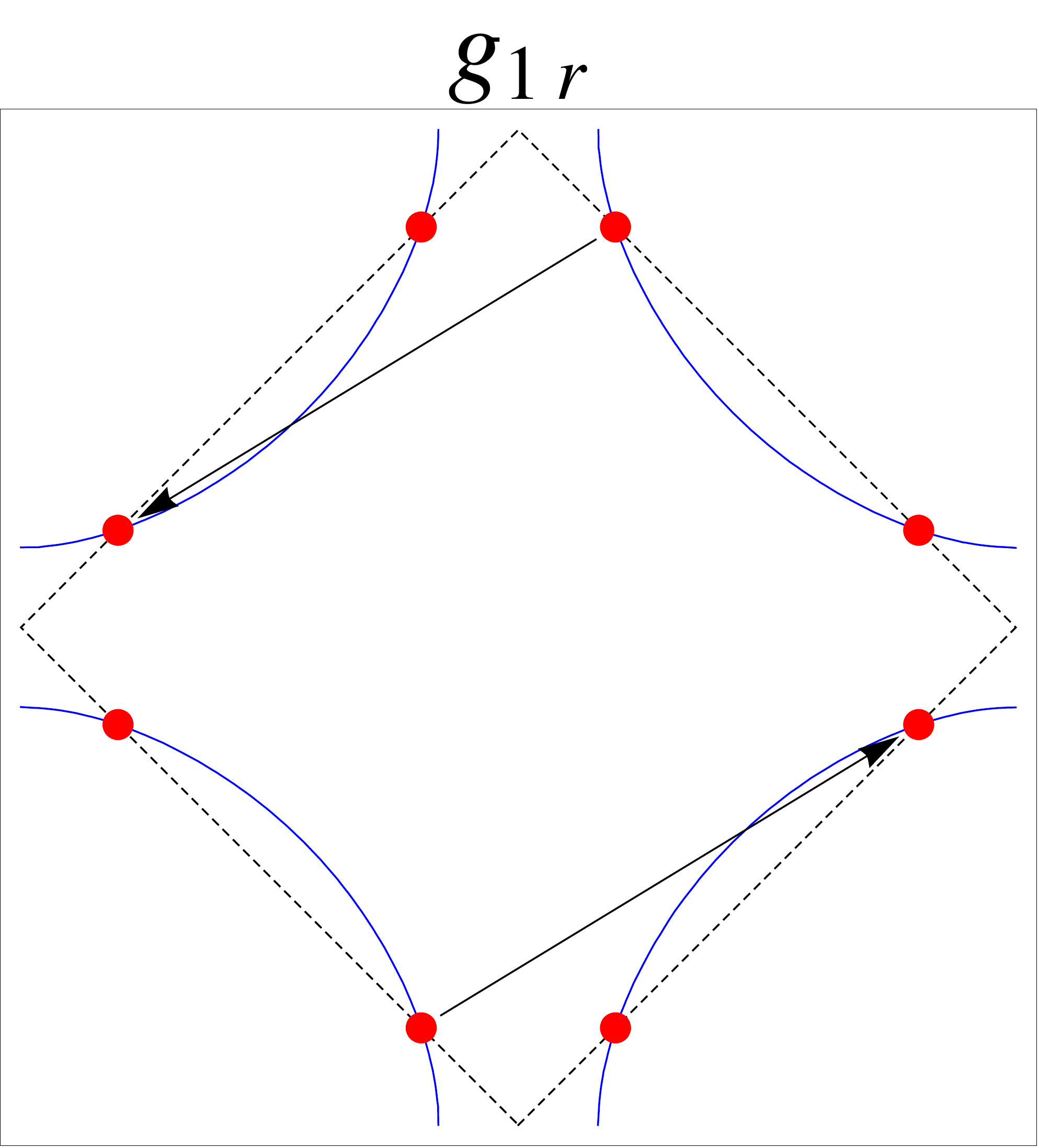}
\includegraphics[height=3.25cm]{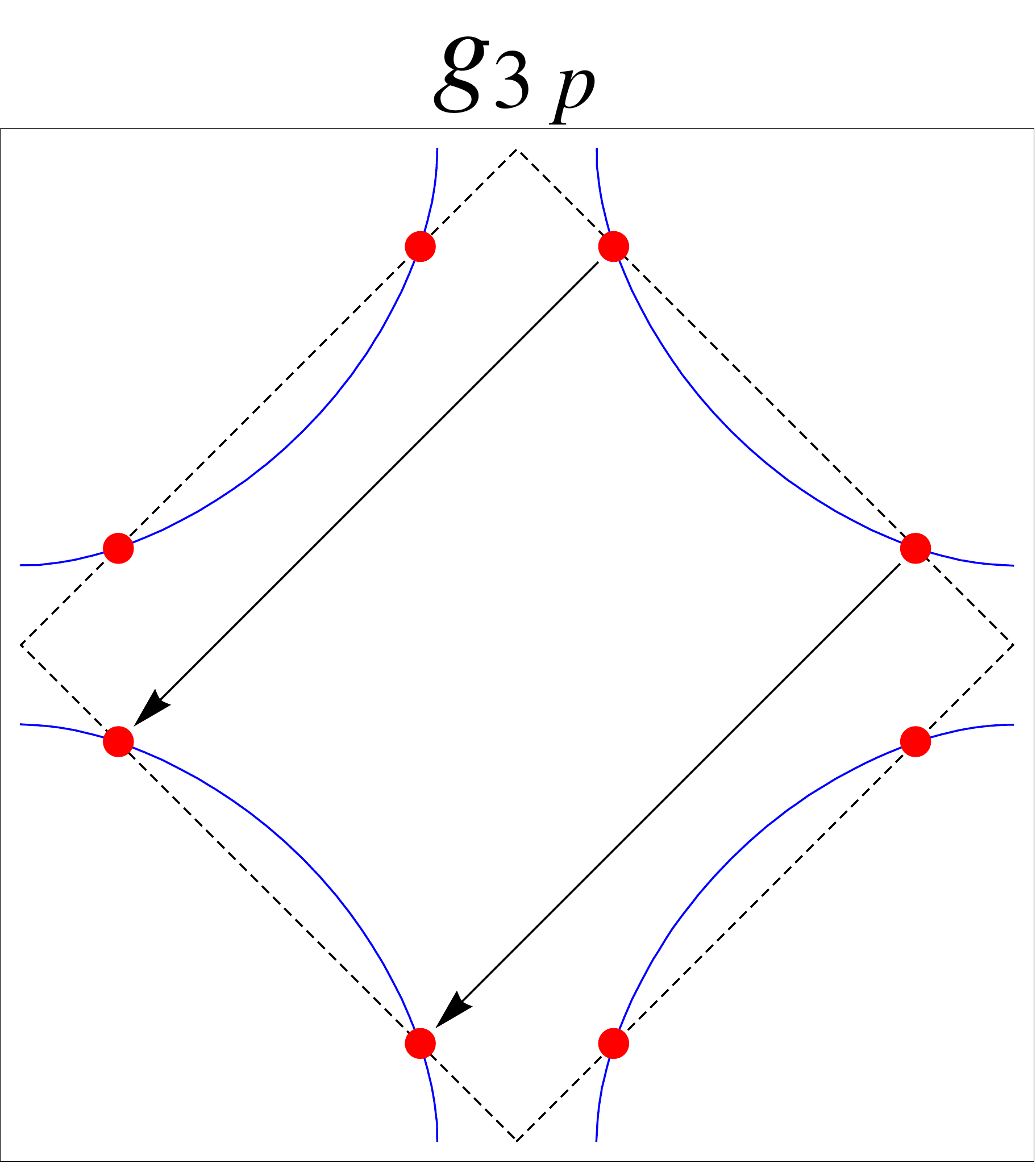}
\includegraphics[height=3.25cm]{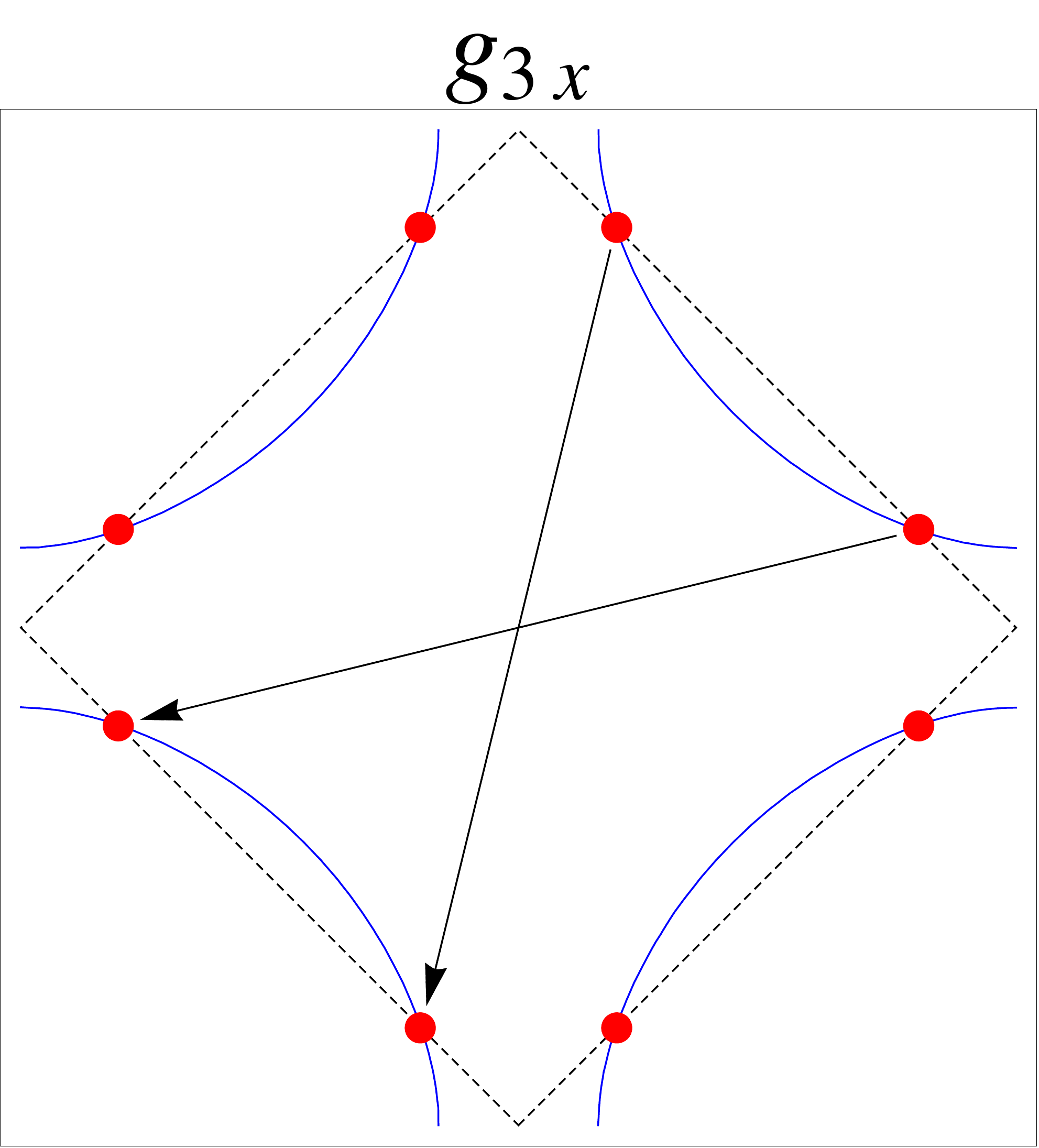}
\includegraphics[height=3.25cm]{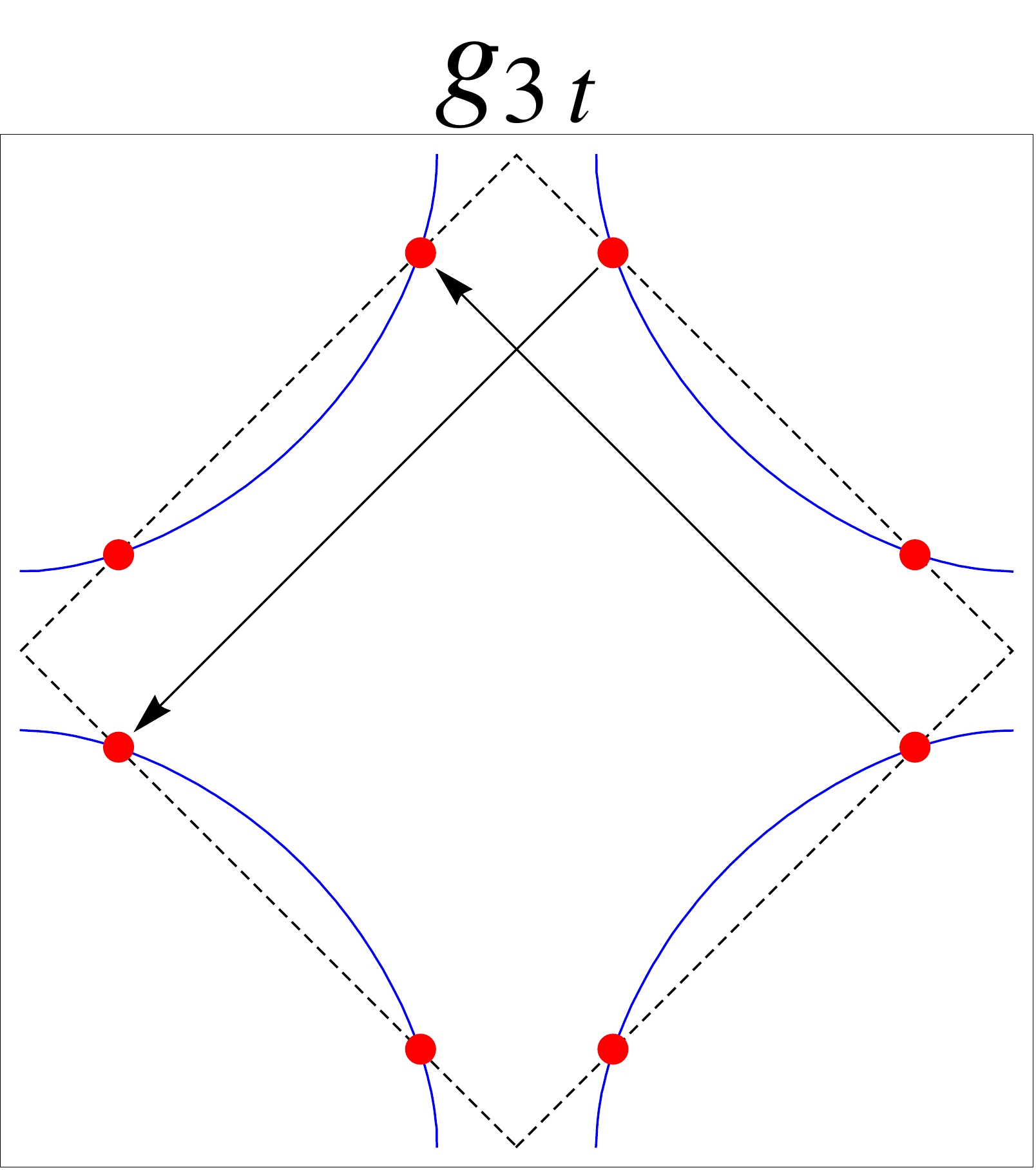}
\includegraphics[height=3.25cm]{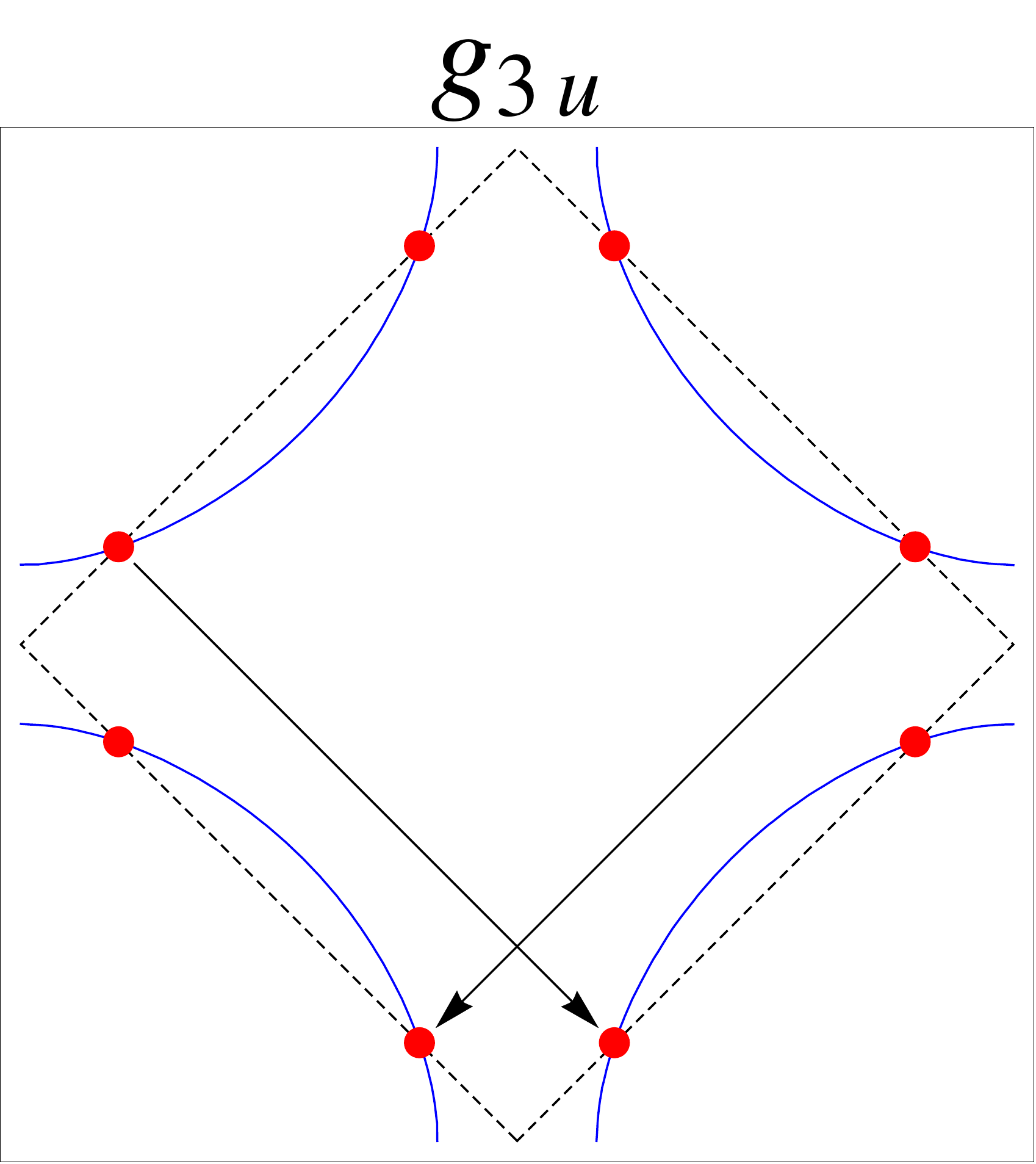}
\includegraphics[height=3.25cm]{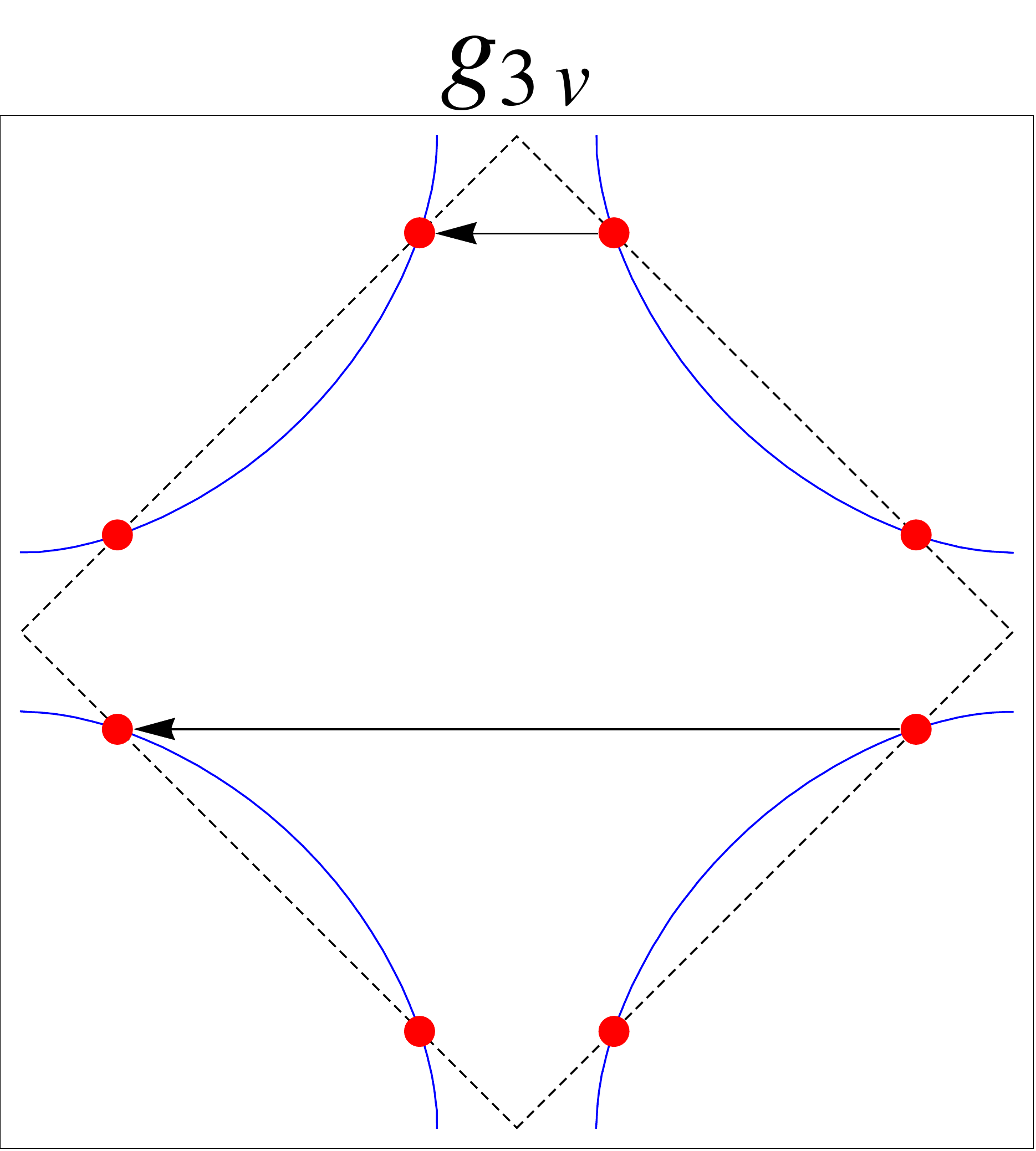}
\includegraphics[height=3.25cm]{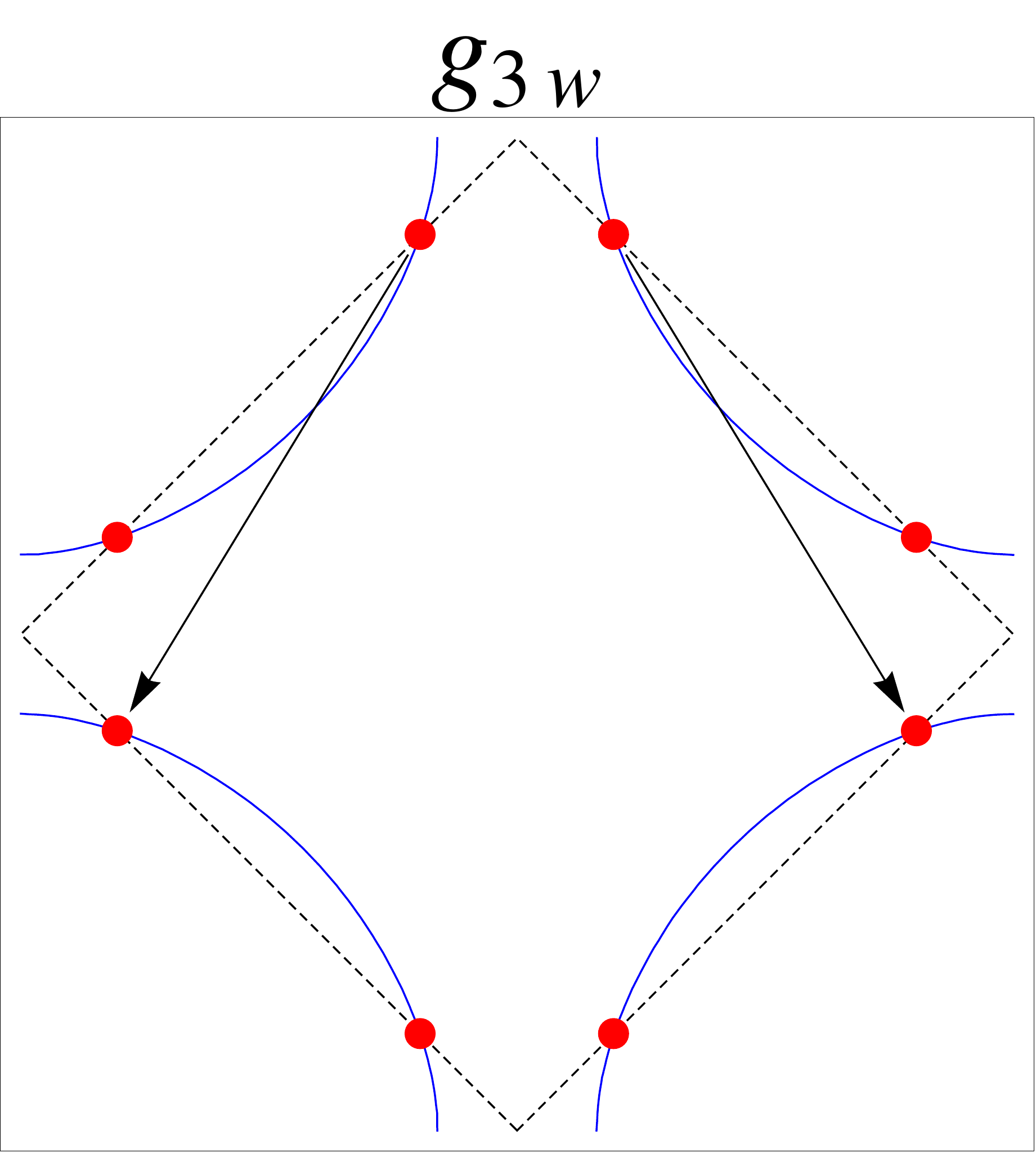}
\caption{(Color online) The relevant couplings for the nested hot spot model. For the $g_{1s}$ coupling, we show two processes which are identical for the interactions considered here.}
\label{fig:couplings}
\end{figure}

Although the number of terms in Eqn. (\ref{eqn:hamiltonian}) appears very large, it is heavily constrained by momentum conservation. For an RG analysis, we only need to consider couplings which are relevant under RG flows. In the nested model, there are $15$ distinct relevant couplings: $n = 1,2,3,1c,2c,1x,2x,1s,1r,3p,3x,3t,3u,3v,3w$, defined in Fig. \ref{fig:couplings}.

\section{Field Theoretic RG}
\label{sec:rg}
\subsection{Couplings}
\label{sec:coup}
In this section, we find the relevant effects of interactions at low energy using standard field-theoretic RG methods.\cite{shankar:rg} We begin by noting that the n-point functions of the bare theory computed at an energy scale $\Lambda$ will depend on $\log ( \Lambda/E_c)$, resulting in logarithmic divergences as we probe physics in the IR limit $\Lambda \rightarrow 0$. Our solution is to define the renormalized couplings as
\begin{equation}\label{eqn:coupdef}
- i N^{-1} g_n(\Lambda) = \langle \psi_{\sigma}(\Lambda/2,\mathbf{k}_{F1n})  \psi_{\sigma'}(\Lambda/2,\mathbf{k}_{F2n}) \overline{\psi}_{\sigma'}(-\Lambda/2,\mathbf{k}_{F3n}) \overline{\psi}_{\sigma}(3\Lambda/2,\mathbf{k}_{F4n}) \rangle
\end{equation}
where the expectation value is computed from the bare theory in presence of the cutoff $E_c$. Here, $N = k_c/(\pi^2 v_F)$ is a conventional rescaling,\cite{solyom,Giamarchi} and for each $g_n$ we label the relevant momenta $\mathbf{k}_{Fjn}$, $j = 1,...,4$ labeling which hot spot each fermion comes from. We can now express all observables in terms of the renormalized, scale-dependent $g_n(\Lambda)$. In addition, since our couplings are now scale-dependent, we can explicitly investigate the leading contributions to the physics at low-energy.

\begin{figure}[h]
\flushleft
\includegraphics[height=3cm]{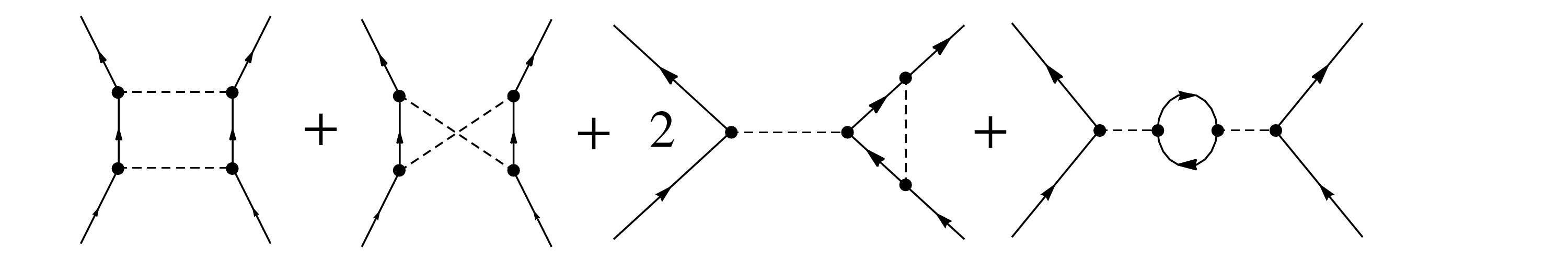}
\caption{Feynman diagrams contributing to the one-loop RG equations. The dashed line denotes a four-Fermi interaction, and the vertices conserve spin.}
\label{fig:oneloop}
\end{figure}

The relevant Feynman diagrams for the one-loop renormalization of the coupling constants are shown in Fig. \ref{fig:oneloop}. The particle-hole and particle-particle bubbles will only contain logarithmic divergences if the momenta in the loop are nested and antiparallel, so the relevant integrals become similar to the one-dimensional case. Once we have the logarithmic dependence of the renormalized couplings, we derive the RG equations which describe their dependence on the energy scale. The RG equations obtained by this method are given in Appendix \ref{appa}.

In a general RG treatment, we must also take into account the renormalization of the Fermi surface, the Fermi velocity, and the field itself. These are all renormalized by the self-energy diagrams. However, it is well-known that the self-energy of four-Fermi interactions is independent of the renormalization scale at one-loop,\cite{shankar:rg} so we should ignore these effects at this order.

\subsection{Pairing Vertex}
\label{sec:linresp}

In this section, we follow the notation of Metlitski and Sachdev.\cite{metlitski:prb10.2} We are interested in the renormalization of the spin singlet, even parity, zero-momentum superconducting order parameter, and there are four distinct order parameters we can form out of the four pairs of hot spots:
\begin{equation}\label{eqn:pairing}
V_{\mu \nu} = \epsilon_{\sigma \sigma'} \left( \psi_{1\sigma} \psi_{5\sigma'} + \mu \psi_{4 \sigma} \psi_{8\sigma'} \right) + \nu \epsilon_{\sigma \sigma'}\left( \psi_{3\sigma} \psi_{7\sigma'} + \mu \psi_{6\sigma} \psi_{2\sigma'} \right).
\end{equation}
Here, the subscripts $i = 1,2,...,8$ denote the hot spots as labelled in Fig. \ref{fig:hotspot}, and the coefficients $\mu = \pm1$, $\nu = \pm1$ determine the transformation properties of $V_{\mu \nu}$ under the discrete symmetries in the Brillouin zone. Explicitly, we define counterclockwise rotation through an angle $\pi/2$ by $R_{\pi/2}$, and reflection symmetry about the $\hat{y}$ axis by $I_{\hat{y}}$, obtaining the relations
\begin{eqnarray}\label{eqn:ordsym}
R_{\pi/2}: \qquad V_{\mu \nu} &\rightarrow& \nu V_{\mu \nu} \\
I_{\hat{y}}: \qquad V_{\mu \nu} &\rightarrow& \mu V_{\mu \nu}.
\end{eqnarray}
We summarize these properties in Table \ref{tab:symm}.
\begin{table}
\begin{tabular}{cr|cc}
\toprule
& & \multicolumn{2}{c}{$\mu$} \\ & &  +1   &  -1   \\
       \hline
\multirow{2}{*}{$\nu \qquad$} &+1  & $s$ & $g$ \\
&-1  & $\quad d_{xy} \quad$ & $d_{x^2 - y^2}$ \\
\toprule
\end{tabular}
\caption{Symmetry properties of the singlet pairing vertex, from Ref.\cite{metlitski:prb10.2}}\label{tab:symm}
\end{table}
We will find that our model gives distinct order parameters for all four symmetry classes.

\begin{figure}[h]
\flushleft
\includegraphics[width=5cm]{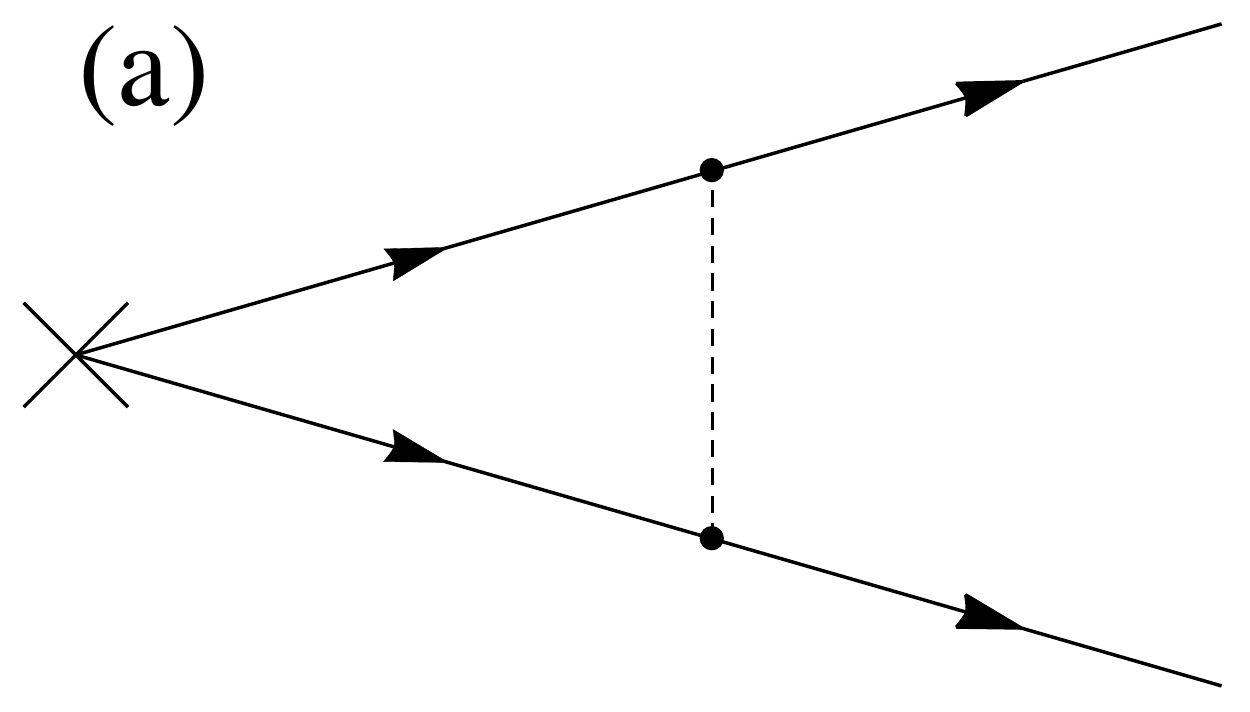}
\includegraphics[width=18cm]{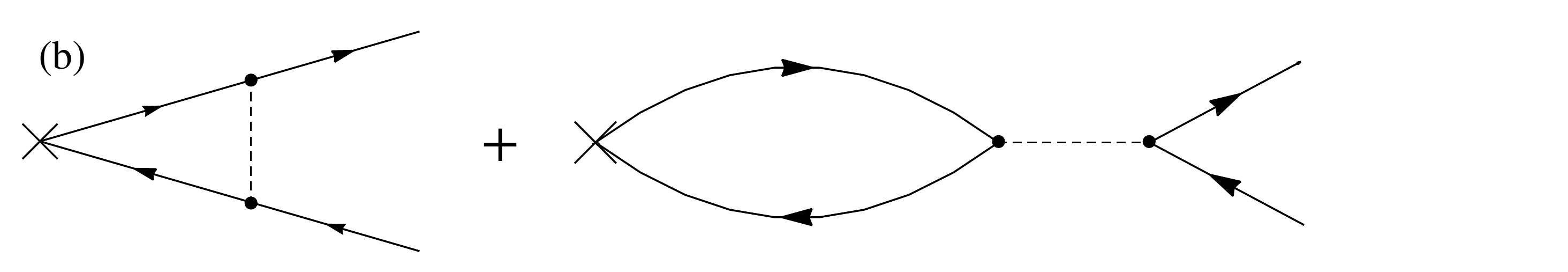}
\caption{The one-loop Feynman diagrams needed to renormalize the (a) singlet pairing susceptibility and (b) the density wave susceptibility. The dashed line denotes a four-Fermi interaction, and the vertices conserve spin.}
\label{fig:vertices}
\end{figure}

We now define our order parameter in terms of the correlation function
\begin{equation}\label{eqn:pairop}
- i \chi^{SSC}_{\mu \nu}(\Lambda) = \epsilon_{\sigma \sigma'}\langle V_{\mu \nu}(\Lambda,\mathbf{k}_F) \overline{\psi}_{1,\sigma}(0,\mathbf{k}_F) \overline{\psi}_{5 \sigma'}(0,\mathbf{k}_F) \rangle
\end{equation}
where we have again used the RG scale $\Lambda$, and it is understood that the Fermi momenta $\mathbf{k}_F$ correspond to the wave vectors associated with the labeled hot spot fermions. The relevant Feynman diagrams up to one-loop are shown in Fig. \ref{fig:vertices}(a). At one-loop our renormalized susceptibilities all satisfy RG equations of the form
\begin{equation}\label{eqn:linrg}
\Lambda \frac{d\chi_{\mu \nu}^{SSC}}{d \Lambda} = \alpha_{\mu \nu}^{SSC} \chi_{\mu \nu}^{SSC}
\end{equation}
which have the solutions
\begin{equation}\label{eqn:linform}
\chi_{\mu \nu}^{SSC}(\Lambda) = \chi_{\mu \nu}(\Lambda_0) \left( \frac{\Lambda}{\Lambda_0} \right)^{\alpha_{\mu \nu}^{SSC}}
\end{equation}
and $\alpha_{\mu \nu}^{SSC}$ is a function of the renormalized couplings from the last section. As we flow to $\Lambda \rightarrow 0$, the susceptibilities will either go to zero ($\alpha > 0$) or diverge ($\alpha < 0$). The exponents for the singlet susceptibilities at one-loop are independent of whether we have $\theta=0$ or $\theta >0$, and are given by
\begin{eqnarray}\label{eqn:crexpssc}
\alpha_{\mu \nu}^{SSC} = \frac{1}{2} \left( g_{1c} + g_{2c} + \mu g_{1x} + \mu g_{2x} + 2 \nu g_{1s} + 2 \mu \nu g_{1r} \right).
\end{eqnarray}

\subsection{Density Vertex}
\label{sec:densvert}

We now consider the possible density order parameters relevant under RG scaling. This proceeds in a similar fashion to the pairing vertex. We define
\begin{eqnarray}
V^{\mathbf{q}}_{\sigma \sigma'}(\Lambda,\mathbf{k}) &=& \overline{\psi}_{\sigma}(\Lambda,\mathbf{k}+\mathbf{q}) \psi_{\sigma'}(\Lambda,\mathbf{k}) \\ \label{eqn:pair}
-i \chi^{\mathbf{q},DW}_{\sigma \sigma'}(\Lambda) &=& \langle V^{\mathbf{q}}_{\sigma \sigma'}(\Lambda,\mathbf{k}_F) \overline{\psi}_{\sigma}(0,\mathbf{k}+\mathbf{q}) \psi_{\sigma'}(0,\mathbf{k}) \rangle
\end{eqnarray}
where $V^{\mathbf{q}}_{\sigma \sigma'}(\mathbf{k})$ is the density vertex for ordering wave vector $\mathbf{q}$, $\chi^{\mathbf{q},DW}_{\sigma \sigma'}$ is the corresponding susceptibility, and there is no sum on the repeated spin indices in Eqn. (11). Once these generalized susceptibilities are computed, one can form the charge and spin susceptibilities as
\begin{eqnarray}
\chi^{\mathbf{q}}_{CDW} &=& \chi^{\mathbf{q},DW}_{\uparrow \uparrow} + \chi^{\mathbf{q},DW}_{\downarrow \downarrow} \\
\chi^{\mathbf{q}}_{SDW} &=& \chi^{\mathbf{q},DW}_{\uparrow \uparrow} - \chi^{\mathbf{q},DW}_{\downarrow \downarrow}.
\end{eqnarray}

With these definitions it is straight-forward to perform the calculations, and the relevant Feynman diagrams are shown in Fig. \ref{fig:vertices}(b). For $\theta = 0$, we find that the only ordering wave vectors with relevant susceptibilities under RG flow are $\mathbf{q} = (\pi,\pi)$ and $\mathbf{q} = (Q_0,Q_0)$ (shown in Fig. \ref{fig:hotspot}), and the critical exponents are
\begin{eqnarray}\label{eqn:crexp}
&\alpha_{CDW}^{(\pi,\pi)} & = \frac{1}{2} \left( 2 g_1 - g_2 + g_3 + 2 g_{1x} - g_{2x} - g_{3x} + 2 g_{3p} + 4 g_{3t} + 4 g_{3u} - 2 g_{3v} - g_{3w} \right) \\
&\alpha_{SDW}^{(\pi,\pi)} & = - \frac{1}{2} \left( g_2 + g_3 + g_{2x} + g_{3x} + 2 g_{3v} + 2 g_{3w} \right) \\
&\alpha_{CDW}^{(Q_0,Q_0)} & = \frac{1}{2} \left( 2 g_{1c} - g_{2c} - g_{3p} + 2 g_{3x} \right) \\
&\alpha_{SDW}^{(Q_0,Q_0)} & = - \frac{1}{2} \left( g_{2c} + g_{3p} \right).
\end{eqnarray}
For $\theta \in (0,\pi/4)$, the $(\pi,\pi)$ susceptibilities are irrelevant while the $(Q_0,Q_0)$ instabilities are unchanged. We note that these equations agree with known results in the 1D limit and the limit of zero chemical potential.\cite{solyom,furukawa98}

In addition to these order parameters, we also investigate density fluctuations at $(Q_0,Q_0)$ with $d$ symmetry. We form this vertex explicitly from the hot spots as
\begin{eqnarray}
\widetilde{V}^{(Q_0,Q_0)}_{\sigma \sigma'} = \overline{\psi}_{3 \sigma}\psi_{7 \sigma'} - \overline{\psi}_{4 \sigma} \psi_{8 \sigma'},
\end{eqnarray}
so the vertex changes sign under $\pi/2$ rotations. The critical exponents for these order parameters are independent of nesting, and given by
\begin{eqnarray}
&\alpha_{\widetilde{CDW}}^{(Q_0,Q_0)} & = \frac{1}{2} \left( 2 g_{1c} - g_{2c} + g_{3p} - 2 g_{3x} \right) \\
&\alpha_{\widetilde{SDW}}^{(Q_0,Q_0)} & = \frac{1}{2} \left( g_{3p} - g_{2c} \right) 
\end{eqnarray}
where the notation $\widetilde{SDW}$ and $\widetilde{CDW}$ distinguishes these from the above exponents. We note that the $\widetilde{CDW}$ order parameter corresponds to the $d$-form factor charge order which has been suggested in explaining the experiments referred to above.

\section{Results}
\label{sec:results}
\subsection{$\theta = 0$}

We now integrate the RG equations and observe the leading divergences in the susceptibilities. These divergences will signal the instability of the free theory to long-range order induced by interactions. We solve the 15 coupled RG equations numerically using a fourth-order Runge-Kutta method. We write the RG scale as $\Lambda = \exp(-l)$, so that the differential operators in Appendix \ref{appa} take the form $\Lambda d/d\Lambda = -d/dl$ where $l$ is our single RK4 step.

We note that the solution to our one-loop RG equations always diverge at some critical step $l_c$, where the interactions flow to strong coupling at some small finite $\Lambda_c$. This divergence is an artifact of the one-loop calculation, and higher order terms will shift this divergence to $\Lambda = 0$.\cite{solyom:73,*solyom:73.1,furukawa98} At the end of our RG flow at $l_c$, the perturbative treatment breaks down, so our results should be interpreted as indicating a tendency to a particular strong coupling fixed point rather than a rigorous determination of the phase diagram.

We now turn to the problem of choosing initial conditions for the 15-dimensional parameter space spanned by our couplings. The solutions to the RG equations are heavily dependent on the initial conditions, so we should use relevant physical models to motivate our choices. In this paper we consider two different interactions. The first is the Hubbard interaction
\begin{eqnarray}\label{eqn:hubbard}
\mathcal{S}_{Hubbard} &=& -U \sum_{\sigma \sigma'}\sum_i \overline{\psi}_{\sigma' }(\mathbf{r}_i) \psi_{\sigma' }(\mathbf{r}_i) \overline{\psi}_{\sigma}(\mathbf{r}_i) \psi_{\sigma}(\mathbf{r}_i) \nonumber \\
&=& -U  \sum_{\sigma \sigma'} \int \left( \prod_{i=1}^4 d^3k_i \right) \bar{\delta}(k_1 + k_2 - k_3 - k_4) \overline{\psi}_{\sigma }(k_4) \overline{\psi}_{\sigma' }(k_3) \psi_{\sigma' }(k_2) \psi_{\sigma}(k_1) .
\end{eqnarray}
Comparing this with equations (\ref{eqn:hamiltonian}) and (\ref{eqn:coupdef}), one can see that our renormalized $g$-ology couplings are all equal, and given by $g_n = \frac{k_c}{\pi^2 v_F}U$. We will also consider a $J$, $V$ interaction, where $J$ represents a Heisenberg exchange coupling and $V$ represents nearest-neighbor repulsion:
\begin{eqnarray}\label{eqn:tjv}
\mathcal{S}_{JV} &=& -J \sum_{\langle ij \rangle} \vec{S}_i \cdot \vec{S}_{j} - V \sum_{\sigma \sigma'}\sum_{\langle ij \rangle} n_{i\sigma} n_{j\sigma'} \nonumber \\
&=& -\sum_{\sigma \sigma'}  \int \left( \prod_{i=1}^4 d^3k_i \right) \bar{\delta}(k_1 + k_2 - k_3 - k_4) \mathcal{G}(J,V,\mathbf{k}_i) \overline{\psi}_{\sigma }(k_4) \overline{\psi}_{\sigma' }(k_3) \psi_{\sigma' }(k_2) \psi_{\sigma}(k_1)
\end{eqnarray}
where $n_{i\sigma} = \overline{\psi}_{\sigma}(\mathbf{r}_i)\psi_{\sigma}(\mathbf{r}_i)$, $\vec{S}_i = \frac{1}{2} \overline{\psi}_{\sigma}(\mathbf{r}_i) \vec{\tau}_{\sigma \sigma'} \psi_{\sigma'}(\mathbf{r}_i)$, and
\begin{eqnarray}\label{eqn:tjvcoup}
\mathcal{G}(J,V,\mathbf{k}_i) &=& -J/2 \left( \cos(k_{1x} - k_{3x}) + \cos(k_{1y} - k_{3y})  \right) \nonumber \\
&+& \left( V - J/4 \right) \left( \cos(k_{2x} - k_{3x}) + \cos(k_{2y} - k_{3y}) \right).
\end{eqnarray}
Once again, we can relate the initial conditions on $J$ and $V$ to initial conditions on the $g$ couplings by expanding equation (\ref{eqn:tjvcoup}) around each hot spot and matching with (\ref{eqn:hamiltonian}) term by term. In stating our results, it will be more convenient to measure all energies in units of $\pi^2v_F/k_c$, defining $\tilde{J}$ and $\tilde{V}$ to be scaled by this quantity. In expanding (\ref{eqn:tjvcoup}) at the hot spots, we take $t'/t = -0.3$ and $\mu/t = -1.1$.
\begin{figure}[h]
\includegraphics[height=4.5cm]{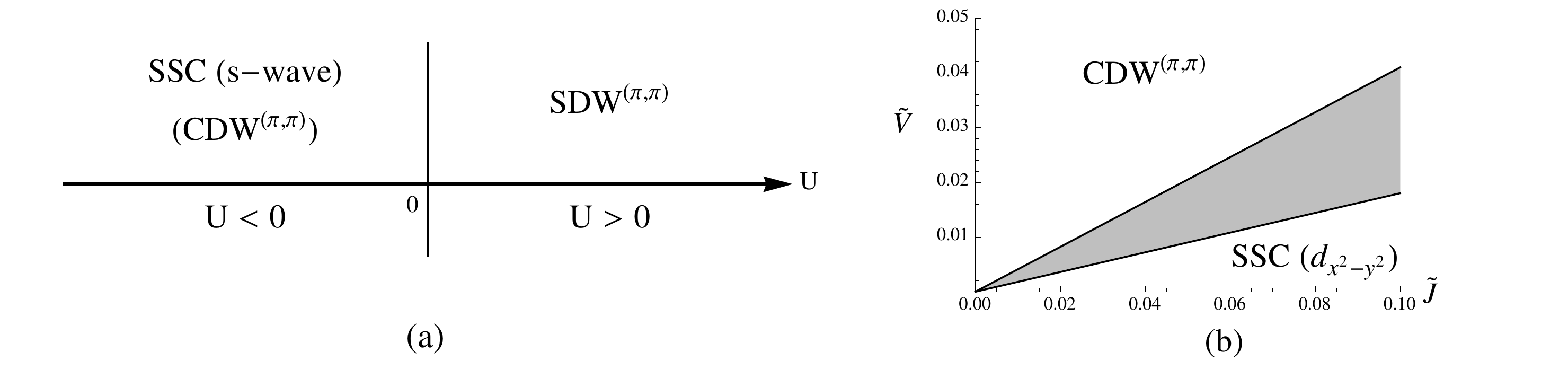}
\caption{Dominant instabilities for the nested hot spot model with (a) Hubbard and (b) $J$, $V$ interactions. The couplings $\tilde{J}$ and $\tilde{V}$ are scaled by $k_c/\pi^2v_F$. The shaded region in (b) displays competing order between SSC ($d_{x^2-y^2}$), SDW$^{(\pi,\pi)}$, and $\widetilde{CDW}^{(Q_0,Q_0)}$.} 
\label{fig:phases1}
\end{figure}

The leading instabilities for the nested hot spot Hubbard model are shown in Fig. \ref{fig:phases1}(a). The results are only dependent on the sign of $U$. We find that the leading instability for $U>0$ is to N\'eel order, while for $U<0$ we find $s$-wave singlet superconductivity dominant followed by commensurate CDW$^{(\pi,\pi)}$ order. These results resemble those obtained from numerical studies on the half-filled Hubbard model, which find N\'eel order for $U>0$ and competing $s$-wave superconductivity and CDW$^{(\pi,\pi)}$ order for $U<0$.\cite{hirsch:prb85} These similarities might suggest that the nested hot spot Hubbard model is related to the half-filled Hubbard model.

The results for the nested hot spot $J$, $V$ model are shown in Fig. \ref{fig:phases1}(b). For large exchange coupling $J$, the model exhibits enhanced singlet $d_{x^2-y^2}$ pairing, but for large $V$ the system crosses over to a commensurate CDW$^{(\pi,\pi)}$ state. However, these phases are separated by a region of competing order, where the $d_{x^2-y^2}$ pairing, commensurate SDW$^{(\pi,\pi)}$, and $\widetilde{CDW}^{(Q_0,Q_0)}$ are all strongly divergent. The nature of the ground state of this region is likely out of the scope of the present calculation, and would require higher-order calculations to investigate. 

\subsection{$\theta > 0$}
\label{sec:angle}

\begin{figure}[h]
\includegraphics[width=7.5cm]{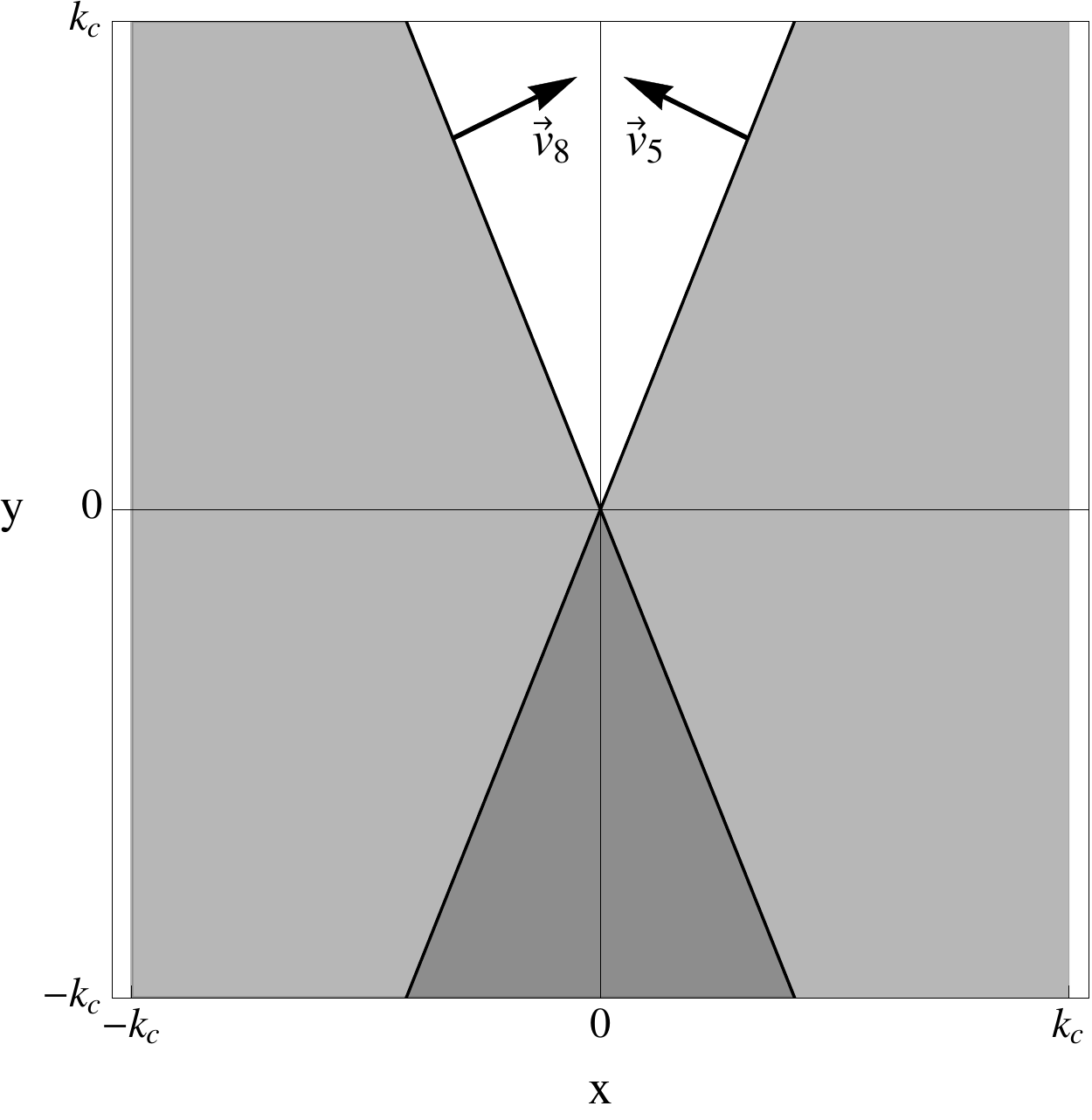}
\caption{Region of integration between two hot spots separated by the wave vector $(\pi,\pi)$. The axes and hot spot labels agree with Fig. \ref{fig:hotspot}. The lightly shaded regions denote singly occupied momenta while the dark shaded region is doubly occupied.} 
\label{fig:integ}
\end{figure}

We now review the case where the hot spots are not nested, $\theta \in (0,\pi/4)$ (see Fig. \ref{fig:hotspot}), first considered in 
Ref.~\onlinecite{furukawa98}. In this limit, only the hot spots connected by the incommensurate wave vector $(Q_0,Q_0)$ remain nested, while the other hot spot dispersions are no longer strictly parallel or perpendicular. However, the momentum integration can still be done exactly. As an example, we consider the particle-hole bubble for two hot spots separated by the momentum $(\pi,\pi)$ (dispersions shown in Fig. \ref{fig:integ}). The particle-hole bubble is given by
\begin{eqnarray}\label{eqn:bubble}
\Delta_{ph} &=& \frac{- k_c}{ 2 \pi^2 v_F \cos \theta} + \frac{k_c}{4 \pi^2 v_F \cos \theta} \log \left| \frac{\Lambda^2 - 4 v_F^2 k_c^2 \sin^2 \theta}{\Lambda^2 - 4 v_F^2 k_c^2 \cos^2 \theta} \right| \nonumber \\
&~&~~~~~~+ \frac{\Lambda}{8 \pi^2 v_F^2 \sin \theta \cos \theta} \log \left| \frac{\Lambda + 2 v_F k_c \sin \theta}{\Lambda - 2 v_F k_c \sin \theta} \right| \nonumber \\
&  & \nonumber \\
& \xrightarrow{\Lambda \rightarrow 0}& \,\, \mbox{const.} + \frac{k_c}{2 \pi^2 v_F cos \theta} \log\left( \tan \theta \right).
\end{eqnarray}

We see that for any finite angle, this channel will not contribute any logarithmic divergences in the IR. Similar results hold for all other non-nested loop diagrams. Furthermore, while we expect the angle $\theta$ to be renormalized in the presence of interactions, possibly realizing divergences in this channel, this renormalization will not occur until two-loop. Therefore, the finite $\theta$ case has very different properties than the nested hot spot case considered above, and the limit $\theta \rightarrow 0$ should not be taken carelessly.

\begin{figure}[h]
\includegraphics[height=4.5cm]{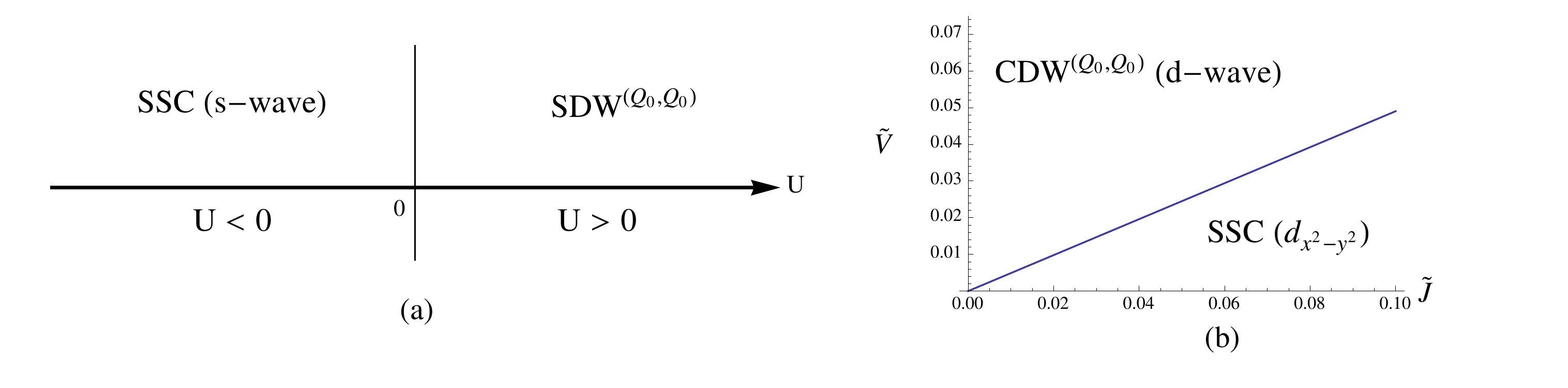}
\caption{Dominant instabilities for the $\theta \in (0,\pi/4)$ hot spot model with (a) Hubbard and (b) $J$, $V$ interactions. The couplings $\tilde{J}$ and $\tilde{V}$ are scaled by $k_c/\pi^2v_F$.}
\label{fig:phases2}
\end{figure}
The dominant instabilities in the non-nested hot spot model for Hubbard and $J$, $V$ interactions are shown in Fig. \ref{fig:phases2}. We find the incommensurate SDW order first obtained by Furukawa and Rice\cite{furukawa98} to be dominant for $U>0$, and we find $s$-wave pairing dominant for $U<0$. For $J$, $V$ interactions, we again find enhanced $d_{x^2-y^2}$ pairing for large $J$, but this pairing is suppressed by the nearest-neighbor interaction $V$. For large $V$ the system has a dominant incommensurate $d$-form factor charge order, and in an intermediate region $V \sim J/2$ there is a competition between the two orders. This last result has potential relevancy to the underdoped cuprates, where competition between singlet pairing and charge order has been proposed in explaining recent experiments, as noted in Section~\ref{sec:intro}.
These results are similar to those obtained in 
unrestricted Hartree-Fock computations on the full $J$, $V$ lattice model in Ref.~\onlinecite{sau:14}.

\section{Charge ordering at $(Q_0,0)$}
\label{sec:coq}

While our current model appears to exhibit a $d$-form factor charge order instability for the wave vector $(Q_0,Q_0)$, recent experimental results have suggested order at the wave vector $(Q_0,0)$. We note the possibility that our low-energy field theoretic model does not give the correct wave vector due to microscopic details of the interaction, and that we require specific information about the high-energy modes to obtain the correct instability. In this section, we study that properties of the particle-hole vertex for the wave vector $(Q_0,0)$. While this vertex is not relevant under RG scaling, we can ask how it is is enhanced within our model. For simplicity, we only consider $J$, $V$ interactions in this section. 

We define the density vertex similarly to the presentation in Sec. \ref{sec:rg}. We take our ordering wave vector to be $(Q_0,0)$ and consider the operator
\begin{equation}\label{eqn:cbovert}
V_{\mu \nu}^{(Q_0,0)} = \frac{1}{2}\sum_{\sigma} \left( \overline{\psi}_{6\sigma} \psi_{1\sigma} + \mu \overline{\psi}_{5\sigma}\psi_{2\sigma} + \nu \overline{\psi}_{3\sigma}\psi_{4\sigma} + \mu \nu \overline{\psi}_{8\sigma}\psi_{7\sigma}\right)
\end{equation}
with the same hot spot labels as before, and $\mu,\nu=\pm1$ label discrete symmetries in the Brillouin zone. Defining $I_{\hat{k}_x}$ to be reflection about the $\hat{k}_x = (\hat{x}-\hat{y})/\sqrt{2}$ axis and $T_{(\pi,\pi)}$ to be translation of the hot spots by a wave vector $(\pi,\pi)$, the transformation properties of the vertex are given by
\begin{eqnarray}
I_{\hat{k}_x}: V_{\mu \nu}^{(Q_0,0)} \rightarrow \mu V_{\mu \nu}^{(Q_0,0)} \\
T_{(\pi,\pi)}: V_{\mu \nu}^{(Q_0,0)} \rightarrow \nu V_{\mu \nu}^{(Q_0,0)}.
\end{eqnarray}
We summarize these properties in Table \ref{tab:cbo}. In the following, it will be more convenient to define the four-dimensional vectors $V_{\alpha}$, $s_{\alpha}$, $d_{\alpha}$, $p_{\alpha}$ and $f_{\alpha}$ as
\begin{table}
\begin{tabular}{cr|cc}
\toprule
& & \multicolumn{2}{c}{$\mu$} \\ & &  +1   &  -1   \\
       \hline
\multirow{2}{*}{$\nu \qquad$} &+1  & $s$ & $f_x$ \\
&-1  & $\quad d_{x^2-y^2} \quad$ & $p_x$ \\
\toprule
\end{tabular}
\caption{Symmetry properties of the $(Q_0,0)$ density vertex.}\label{tab:cbo}
\end{table}
\begin{equation*}
\mathbf{V} = \sum_{\sigma}\begin{pmatrix} \overline{\psi}_{6\sigma} \psi_{1\sigma} \\ \overline{\psi}_{5\sigma}\psi_{2\sigma} \\ \overline{\psi}_{3\sigma}\psi_{4\sigma} \\ \overline{\psi}_{8\sigma}\psi_{7\sigma} \end{pmatrix},
\end{equation*}
\begin{equation}
\mathbf{s} = \frac{1}{2}\begin{pmatrix} 1 \\ 1 \\ 1 \\ 1  \end{pmatrix}, \quad
\mathbf{p} = \frac{1}{2}\begin{pmatrix} 1 \\ -1 \\ -1 \\ 1  \end{pmatrix},\quad
\mathbf{d} = \frac{1}{2}\begin{pmatrix} 1 \\ 1 \\ -1 \\ -1  \end{pmatrix},\quad
\mathbf{f} = \frac{1}{2}\begin{pmatrix} 1 \\ -1 \\ 1 \\ -1  \end{pmatrix}.
\end{equation}
With these definitions, along with Eqn. (\ref{eqn:cbovert}), we can now note the equalities $\mathbf{V} \cdot \mathbf{s} = V_{11}^{(Q_0,0)}$, $\mathbf{V} \cdot \mathbf{p} = V_{-1-1}^{(Q_0,0)}$, $\mathbf{V} \cdot \mathbf{d} = V_{1-1}^{(Q_0,0)}$, and $\mathbf{V} \cdot \mathbf{f} = V_{-11}^{(Q_0,0)}$, justifying this notation in light of Table \ref{tab:cbo}. 

\begin{figure}
\flushleft
\includegraphics[width=13cm]{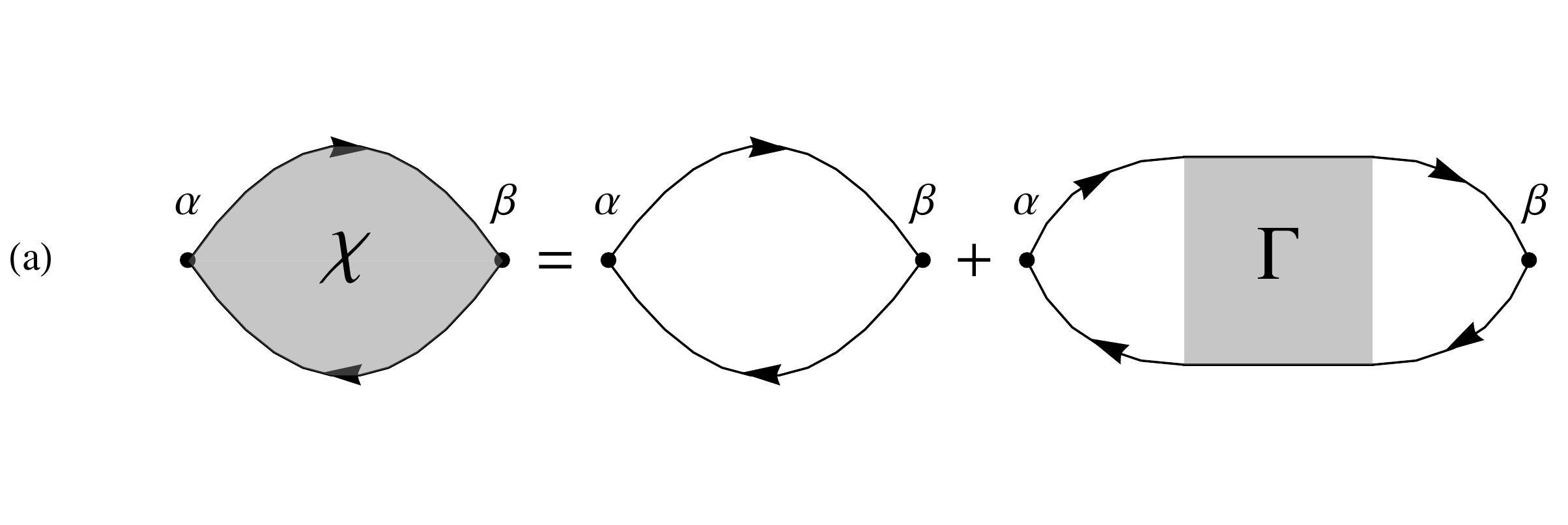}
\includegraphics[width=14cm]{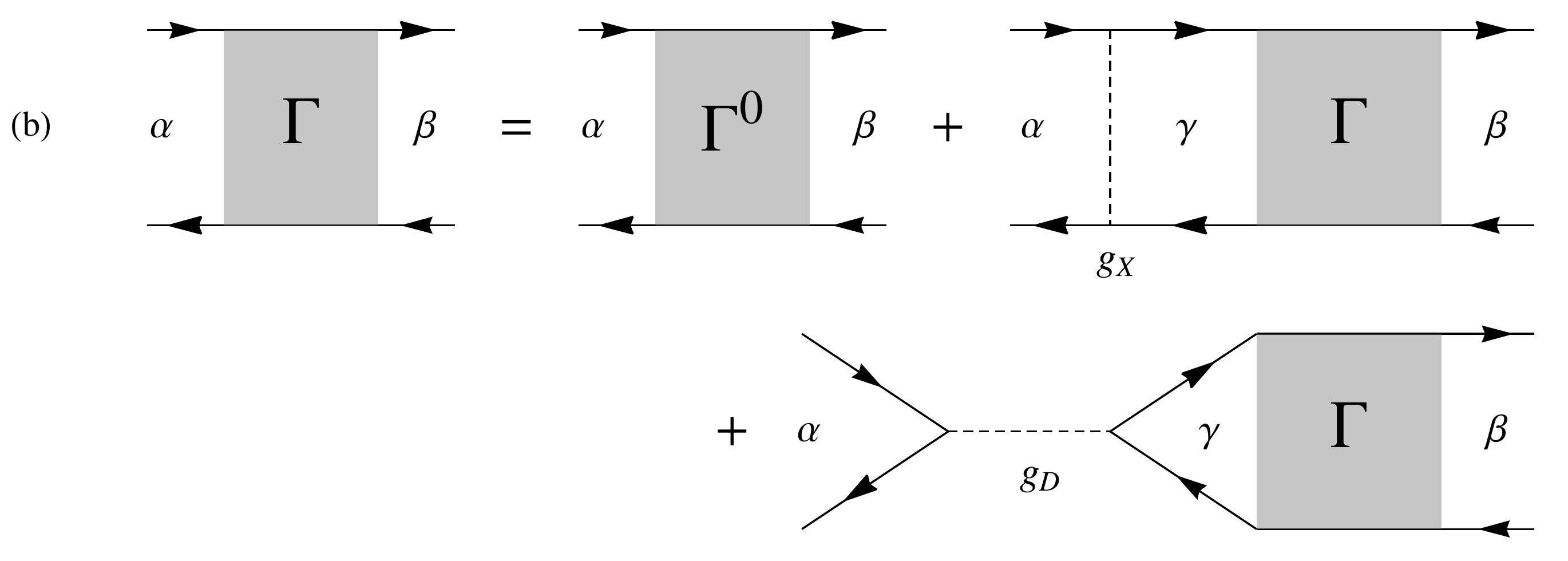}
\caption{(a) Diagrammatic representation of the susceptibility matrix $\chi$ in terms of the effective particle hole interaction $\Gamma$. (b) Diagrammatic representation of the Bethe-Salpeter equation for the particle-hole interaction, see Eqn. (\ref{eqn:bsalp}).}
\label{fig:susc}
\end{figure}

We now define the static charge susceptibility matrix as
\begin{equation}\label{eqn:susc}
-i\chi_{\alpha\beta} = \langle V_{\alpha}(0,\mathbf{k}_F) V_{\beta}(0,\mathbf{k}_F) \rangle
\end{equation}
where the momentum dependence is entirely contained in the indices $\alpha\beta$. We display the Dyson equation for the susceptibility in Fig. \ref{fig:susc}(a). We find that the leading instability is entirely determined by the Bethe-Salpeter equation for the effective particle-hole interaction. From Fig. \ref{fig:susc}(b), we have
\begin{equation}\label{eqn:bsalp}
\Gamma_{\alpha \beta} = \Gamma^0_{\alpha\beta} + \left( g_X \Pi_{\gamma} \right)_{\alpha \gamma} \Gamma_{\gamma \beta} - 2 \left( g_D \Pi_{\gamma} \right)_{\alpha \gamma} \Gamma_{\gamma \beta}.
\end{equation}
Here, $\Gamma^0$ is the tree-level result for the effective interaction, $\left(g_{D,X}\right)_{\alpha\gamma}$ is shorthand for the $g$-ology couplings between $\alpha$ and $\gamma$ in the direct and exchange interaction channels, and $\Pi_{\gamma}$ refers to the loop integral over the particle-hole pair $\gamma$ corresponding to the components of $\mathbf{V}$. By rotational symmetry, we always have $\Pi_1 = \Pi_2$ and $\Pi_3 = \Pi_4$, and all four are the same for $\theta = 0$.

The computation of the particle-hole pairs $\Pi_{\gamma}$ was briefly described in Section \ref{sec:angle}. More explicitly, these factors are given by
\begin{equation}
\Pi_{1(3)} = \lim_{\omega \to 0} \int \frac{d \alpha}{2 \pi} \int \frac{d^2 k}{(2 \pi)^2} \frac{1}{\alpha - \epsilon_{\mathbf{k}1(4)} + i 0^+ \text{sgn}(\epsilon_{\mathbf{k}1(4)})}\frac{1}{\alpha + \omega - \epsilon_{\mathbf{k}6(3)} + i 0^+ \text{sgn}(\epsilon_{\mathbf{k}6(3)})}.
\end{equation} 
Here, $\epsilon_{\mathbf{k}i}$ is the dispersion at the $i$th hot spot, and we take $\omega$ to zero in light of our definition of susceptibility in Eqn. \ref{eqn:susc}. The frequency integration will restrict the momentum integration over singly-occupied momenta, and we still restrict $x,y \in (-k_c,k_c)$ (compare to Fig. \ref{fig:integ}, which shows the particle-hole bubble for hot spots separated by $(\pi,\pi)$). The integrals are exactly evaluated as
\begin{eqnarray}
\Pi_1 &=& \frac{N}{\cos \theta - \sin \theta} \log \left( \frac{2}{1 + \tan \theta} \right) \\
\Pi_3 &=& \frac{N}{\cos \theta + \sin \theta} \log \left( \frac{2}{1 - \tan \theta} \right)
\end{eqnarray}
where $N = k_c/\pi^2v_F$ as above. As required, these coincide for $\theta = 0$, and $\Pi_3$ is logarithmically divergent in the limit $\theta \rightarrow \pi/4$ where the $V_3$ and $V_4$ channels exhibit perfect nesting.

Rearranging eqn. (\ref{eqn:bsalp}), and noting the relation between the static charge susceptibility and the effective interaction, we find that this channel will exhibit an instability whenever there is a zero eigenvalue of the 4x4 matrix
\begin{equation}
M_{\alpha\beta} = \delta_{\alpha\beta} + \left(2 g_D \Gamma_{\beta} - g_X \Gamma_{\beta} \right)_{\alpha\beta}.
\end{equation}
For $\theta \neq 0$, the relation $\Pi_1 \neq \Pi_2$ implies that the eigenvectors will not be symmetric under $T_{(\pi,\pi)}$ defined above. As a result, the eigenvectors will not be in one of the irreducible forms in Table \ref{tab:cbo}, but the exact symmetry under $I_{\hat{k}_x}$ requires the eigenvalues to be a linear combination of either $\mathbf{s}$ and $\mathbf{d}$ or $\mathbf{p}$ and $\mathbf{f}$. 

The spectrum of $M_{\alpha\beta}$ can be computed exactly within our model. Specializing to the values $t'/t = .3$ and $\mu/t=-1.1$, we consider the parameter region $0 \leq \tilde{J},\tilde{V} \leq 1$, $\theta \in (0,\pi/4)$. For this range of parameters, there is only a single instability in the susceptibility. Defining the polarization $\mathbf{P}$ to be the normalized eigenvector corresponding to the zero eigenvalue of $M$, we find that the polarization is always primarily $d$-wave, with a very small $s$-wave component. Explicitly, if we write
\begin{equation}
\mathbf{P} = s \mathbf{s} + d \mathbf{d},
\end{equation}
then within the entire parameter range considered above the coefficients satisfy $|d| \gtrsim 0.95$ and $|s| \lesssim .31$. These inequalities are saturated for larger $\theta$, while for smaller angles the $d$-wave becomes even more dominant. We conclude that this model contains a charge instability at the Hartree-Fock level with a wave vector $(Q_0,0)$ which is almost entirely $d$-wave. Related results appear in 
Ref.~\onlinecite{DCSS14}.

\section{Conclusions}
\label{sec:conclusions}
We considered a field-theoretic hot spot model relevant to the phenomenology of the cuprates, and studied the relevant order parameters under RG flow. While our results are largely qualitative, we found enhancement of $d$-wave pairing due to exchange interactions, and when the FS is not nested this order competes with an incommensurate $d$-form factor charge order. While this charge order enhancement is in the $(1,1)$ direction, differing from the experimentally measured value, we investigate the properties of the charge instability in the relevant $(1,0),(0,1)$ directions at the Hartree-Fock level. We find that the dominant density wave instability has a dominant $d$-form factor, 
consistent with recent STM observations.\cite{fujita:14}

The major difficulties with this method are the presence of strong coupling and the simplifying assumptions of the model. Since our perturbative RG approach always flows to strong coupling, signaling a breakdown of perturbation theory, we cannot trust our results beyond the tendency of certain instabilities to form. While these calculations can be an excellent guide to the nature of the physical strong-coupling fixed point, any quantitative results will not be accurate. In spite of these problems, even obtaining qualitative results for strongly interacting systems is an important step for understanding these materials. Further work along the lines of this model could investigate the effects of Fermi surface curvature in the dispersion by including quadratic terms. 

\begin{acknowledgements}
We acknowledge discussions with H.~Freire on the interpretation of Ref.~\onlinecite{freire:npb}. 
This material is based upon work supported by the National Science Foundation Graduate Research Fellowship Program (S.W.) under Grant No. DGE-1144152. This research
was supported by the NSF under Grant DMR-1103860, the Templeton foundation, and MURI grant W911NF-14-1-0003 from ARO.
Research at Perimeter Institute is supported by the Government of Canada through Industry Canada and by the Province of Ontario through the Ministry of Economic Development \& Innovation.
\end{acknowledgements}

\appendix
\section{RG Equations}
\label{appa}

Below are the renormalization group equations at one-loop. These couplings are defined in equation (\ref{eqn:coupdef}).
\begin{eqnarray}
\Lambda \frac{d g_1}{d \Lambda} &=& g_1^2 + (g_{1x} - g_{2x}) g_{1x} + (g_{3p} - g_{3x}) g_{3p} + 2 (g_{3t} - g_{3v}) g_{3t} + 2 (g_{3u} - g_{3w}) g_{3u} \\
\Lambda \frac{d g_2}{d \Lambda} &=& \frac{1}{2} \left( g_1^2 - g_3^2 - g_{2x}^2 - g_{3x}^2 \right) -  g_{3v}^2 - g_{3w}^2 \\
\Lambda \frac{d g_3}{d \Lambda} &=& (g_1 - 2 g_2) g_3 + (2 g_{3p} - g_{3x}) g_{1x} - (g_{3p} + g_{3x}) g_{2x} + 4 g_{3t} g_{3u} \nonumber \\
&-& 2 g_{3t} g_{3w} - 2 g_{3u} g_{3v} - 2 g_{3v} g_{3w} \\
\Lambda \frac{d g_{1c}}{d \Lambda} &=& g_{1c}^2 + g_{1x} g_{2x} + g_{1s}^2 + g_{1r}^2 + g_{3x}^2 - g_{3p} g_{3x} \\
\Lambda \frac{d g_{2c}}{d \Lambda} &=& \frac{1}{2} \left( g_{1c}^2 + g_{1x}^2 + g_{2x}^2 + 2 g_{1s}^2 + 2 g_{1r}^2 - g_{3p}^2 \right) \\
\Lambda \frac{d g_{1x}}{d \Lambda} &=& g_{1c} g_{2x} + g_{1x} g_{2c} + 2 g_{1s} g_{1r} +(2 g_{1x} - g_{2x}) g_1- g_{1x} g_2+(g_{3p} - g_{3x}) g_3 \nonumber \\
&+&  4 g_{3t} g_{3u} - 2 g_{3t} g_{3w} - 2 g_{3u} g_{3v} \\
\Lambda \frac{d g_{2x}}{d \Lambda} &=& g_{1x} g_{1c} + (g_{2c} - g_2) g_{2x} - g_3 g_{3x} + 2 g_{1s} g_{1r} - g_{3v} g_{3w} \\
\Lambda \frac{d g_{1s}}{d \Lambda} &=& \left( g_{1c} + g_{2c} \right) g_{1s} + \left( g_{1x} + g_{2x} \right) g_{1r} \\
\Lambda \frac{d g_{1r}}{d \Lambda} &=& \left( g_{1c} + g_{2c} \right) g_{1r} + \left( g_{1x} + g_{2x} \right) g_{1s} \\
\Lambda \frac{d g_{3p}}{d \Lambda} &=& \left( 2 g_1 - g_{2c} - g_2 \right) g_{3p} + \left( g_{1x} - g_{2x} \right) g_3 - g_1 g_{3x} \nonumber \\
&+& 2 (g_{3t} - g_{3v}) g_{3t} + 2 (g_{3u} - g_{3w}) g_{3u} \\
\Lambda \frac{d g_{3x}}{d \Lambda} &=& \left( 2 g_{1c} - g_{2c} - g_2 \right) g_{3x} - g_{1c} g_{3p} - g_3 g_{2x} - g_{3v}^2 - g_{3w}^2 \\
\Lambda \frac{d g_{3t}}{d \Lambda} &=& \left( 2 g_1 - g_2 + 2 g_{3p} - g_{3x} \right) g_{3t} + \left( g_3 + 2 g_{1x} - g_{2x} \right) g_{3u} \nonumber \\
&-& \left( g_1 + g_{3p} \right) g_{3v} - \left( g_3 + g_{1x} \right) g_{3w} \\
\Lambda \frac{d g_{3u}}{d \Lambda} &=& \left( 2 g_1 - g_2 + 2 g_{3p} - g_{3x} \right) g_{3u} + \left( g_3 + 2 g_{1x} - g_{2x} \right) g_{3t} \nonumber \\
&-& \left( g_1 + g_{3p} \right) g_{3w} - \left( g_3 + g_{1x} \right) g_{3v} \\
\Lambda \frac{d g_{3v}}{d \Lambda} &=& - \left( g_2 + g_{3x} \right) g_{3v} - \left( g_3 + g_{2x} \right) g_{3w} \\
\Lambda \frac{d g_{3w}}{d \Lambda} &=& - \left( g_2 + g_{3x} \right) g_{3w} - \left( g_3 + g_{2x} \right) g_{3v}
\end{eqnarray}
As discussed in Section \ref{sec:angle}, the RG equations for finite angle can be obtained from these by omitting diagrams with non-nested loops. They were first obtained in Ref.\cite{furukawa98}
\begin{eqnarray}
\Lambda \frac{d g_{1c}}{d \Lambda} &=& g_{1c}^2 + g_{1x} g_{2x} + g_{1s}^2 + g_{1r}^2 + g_{3x}^2 - g_{3p} g_{3x} \\
\Lambda \frac{d g_{2c}}{d \Lambda} &=& \frac{1}{2} \left( g_{1c}^2 + g_{1x}^2 + g_{2x}^2 + 2 g_{1s}^2 + 2 g_{1r}^2 - g_{3p}^2 \right) \\
\Lambda \frac{d g_{1x}}{d \Lambda} &=& g_{1c} g_{2x} + g_{1x} g_{2c} + 2 g_{1s} g_{1r} \\
\Lambda \frac{d g_{2x}}{d \Lambda} &=& g_{1x} g_{1c} + g_{2c} g_{2x} + 2 g_{1s} g_{1r} \\
\Lambda \frac{d g_{1s}}{d \Lambda} &=& \left( g_{1c} + g_{2c} \right) g_{1s} + \left( g_{1x} + g_{2x} \right) g_{1r} \\
\Lambda \frac{d g_{1r}}{d \Lambda} &=& \left( g_{1c} + g_{2c} \right) g_{1r} + \left( g_{1x} + g_{2x} \right) g_{1s} \\
\Lambda \frac{d g_{3p}}{d \Lambda} &=& -g_{2c} g_{3p} \\
\Lambda \frac{d g_{3x}}{d \Lambda} &=& \left( 2 g_{1c} - g_{2c}\right) g_{3x} - g_{1c} g_{3p} 
\end{eqnarray}

\bibliography{whitsitthotspot}

\end{document}